\documentclass[twocolumn,aps,prc,eqsecnum,preprintnumbers,showpacs,
showkeys,nofootinbib,superscriptaddress,fleqn,floatfix,tightenlines, 10pt]
{revtex4-1}
\usepackage{amsmath}
\usepackage{epsfig}
\usepackage{color}
\usepackage{times}
\usepackage{dcolumn}
\everymath{\displaystyle}
\usepackage{bm}
 
\begin{document} 
\title{
Reaction processes of muon-catalyzed fusion in the muonic molecule \boldmath{$dd\mu$}\\
studied with the tractable \boldmath{$T$}-matrix model
}

\author{Qian Wu}
~\email{qwu@nju.edu.cn}
\affiliation{School of Physics, Nanjing University, Nanjing, Jiangsu 210093, China}
\affiliation{Institute of Modern Physics, Chinese Academy of Sciences, Lanzhou 730000, China}

\author{Zhu-Fang Cui}
\email[]{phycui@nju.edu.cn}
\affiliation{School of Physics, Nanjing University, Nanjing, Jiangsu 210093, China}

\author{Masayasu Kamimura}
\email{mkamimura@a.riken.jp}
\affiliation{Nuclear Many-Body Theory  Laboratory, RIKEN Nishina Center, RIKEN, Wako 351-0198, Japan}

 
\begin{abstract}
Muon-catalyzed fusion has recently regained significant attention due to experimental 
and theoretical developments being performed.
The present authors [Phys. Rev. C {\bf 109} 054625 (2024)] proposed the tractable 
\mbox{$T$-matrix} model based on the Lippmann-Schwinger equation to approximate 
the elaborate two- and three-body coupled-channel (CC) calculations 
[Kamimura, Kino, and Yamashita, Phys. Rev. C {\bf 107}, 034607 (2023)]
for the nuclear reaction processes in the muonic molecule  
$dt\mu$, $(dt\mu)_{J=0} \to\!^4{\rm He} + n + \mu + 17.6 \, {\rm MeV}$. 
The $T$-matrix model well reproduced almost all of the results generated by the CC work. 
In the present paper, we apply this model to the nuclear reaction processes 
in the $dd\mu$ molecule,  $(dd\mu)_{J=1} \to\!^3{\rm He} + n + \mu +3.27 \,$ MeV or
$t + p + \mu + 4.03 \,$ MeV, in which the fusion takes place via the $p$-wave 
$d$-$d$ relative motion. Recently, significantly different $p$-wave astrophysical 
$S(E)$ factors of the reaction $d + d \to\!^3{\rm He} + n$ or $t + p$ 
at $E \! \simeq \! 1$ keV to 1 MeV
have been reported experimentally and theoretically by five groups. 
Employing many sets of nuclear interactions that can
reproduce those five cases of $p$-wave $S(E)$ factors,  
we calculate the fusion rate of the $(dd\mu)_{J=1}$ molecule using three kinds of 
methods where results are consistent with each other.
We also derive the $^3{\rm He}$-$\mu$ sticking probability and the absolute values of 
the energy and momentum spectra of the emitted muon.
The violation of charge symmetry in the $p$-wave $d$-$d$ reaction and the $dd\mu$ 
fusion reaction is discussed.
Information on the emitted 2.45-MeV neutrons and \mbox{1 keV-dominant} muons should 
be useful for the application of $dd\mu$ fusion.
\end{abstract}
\maketitle


\section{INTRODUCTION}
A negatively charged muon ($\mu$) injected into the mixture of deuterium ($D$) and 
tritium ($T$) would form a muonic molecule $dt\mu$ with a deuteron ($d$) and a triton ($t$).
Then, the nuclear reaction $dt\mu \to \alpha + n + \mu + 17.6 \,{\rm MeV}$ takes place 
immediately ($\approx \!\!10^{-12}$ s), since the wave functions of $d$ and $t$ overlap 
inside the molecule due to $m_\mu \approx 207$ $m_e$. 
Later on, the free $\mu$ may continue to facilitate another or more \mbox{fusion} reactions. 
This cyclic process is called muon-\mbox{catalyzed fusion ($\mu$CF).}
The $dt\mu$ fusion has attracted particular attention in $\mu$CF as a future energy source.

The $\mu$CF has been dedicatedly investigated since 
\mbox{1947~\cite{Frank1947,Sakharov1948};} cf. review work of 
Refs.~\cite{Breunlich89,Ponomarev90,Bogdanova1988,Froelich92,Nagamine98}.
It has recently attracted again considerable research interest on account of several 
new developments and applications in the experimental and theoretical studies, 
which are briefly reviewed in Ref.~\cite{Kamimura2023}, where Kino, Yamashita and 
one of the present authors (M.K.) comprehensively studied the nuclear reaction processes in
the $dt\mu$ molecule. They employed a three-body coupled-channel (CC) method with the use 
of the nuclear interactions that reproduce the low-energy cross sections of 
the $d + t \to \alpha + n + 17.6 \,{\rm MeV}$ process using a two-body CC method.

Later on, in Ref.~\cite{Wu2024} we proposed a tractable $T$-matrix model to approximate 
the elaborate two- and three-body CC methods for the $dt\mu$ reaction on the basis of 
the Lippmann-Schwinger theory~\cite{Lippmann1950}, and reproduced almost all the results 
of Ref.~\cite{Kamimura2023}.

In the present paper, we apply the $T$-matrix model to the $dd\mu$ reaction.
In the $dd\mu$ molecule, the nuclear reactions,
\begin{eqnarray}
&& d + d \rightarrow \,\! ^3{\rm He} + n  + 3.27 \;\mbox{MeV}, \\
&& d + d \rightarrow t + p  + 4.03 \;\mbox{MeV},
\end{eqnarray}
take place as follows,
\begin{equation*}
 \quad\; \;\;\; (dd\mu^-)_{Jv}  \stackrel{}{\longrightarrow} 
   \begin{cases}
     \:  ^3{\rm He} + n + \mu^-   +3.27 \,\mbox{MeV}, \;\;\: (1.3a)     \\
     \:                             \hskip 4.6cm    \\
     \:  (^3{\rm He}\mu^-) + n    +3.27 \,\mbox{MeV}, \:\, \quad (1.3b) 
   \end{cases}
\end{equation*}
\vskip -0.3cm
\begin{equation*}
\quad \quad \: (dd\mu^-)_{Jv}  \stackrel{}{\longrightarrow} 
   \begin{cases}
     \:  t + p + \mu^- + 4.03 \,\mbox{MeV}, \hskip 0.81cm (1.4c)   \\
     \:  (t\mu^-) + p +4.03 \,\mbox{MeV}, \hskip 0.97cm (1.4b)\\
     \:  (p\mu^-) + t +4.03\,\mbox{MeV}. \hskip 1.07cm (1.4c)
   \end{cases}
\end{equation*}
Namely, fusion occurs in $p$-wave $d$-$d$ relative state with the total angular momentum 
$J\!=\!1$ and spin $S\!=\!1$, because the Pauli principle between the two identical 
bosons prevents de-excitation to the $s$-wave states with $J=S=0$, apart from small 
relativistic effects.

After the fusion takes place, part of the emitted muons stick to $^3{\rm He}$ as 
in Eq.~(1.3b) (much less to $t$ and $p$) with a probability of 
$\approx$13\%~\cite{Balin1984,Bogdanova1985}, the percentage of reaction Eq. (1.3b) in 
the whole Eq.~(1.3). 
This reduces the muon cycling rate down to a level much lower than the scientific 
break-even, and therefore, the $dd\mu$ fusion cannot be utilized alone as an energy source.
However, very recently an interesting use of 
the precisely `2.45 MeV' neutron in the reaction (1.3a) has been proposed by 
Iiyoshi {\it et al.}~\cite{Iiyoshi2023}; it is a \mbox{thorium (Th)} \mbox{subcritical} 
reactor  activated and controlled by 
the $d$-$d$ $\mu$CF,  
which has the potential to be safer, smaller, and generate less radioactive waste 
compared to traditional energy sources over the next few decades.

Since the $dd\mu$ fusion does not need $t$ as a source, the whole $d$-$d$ $\mu$CF mechanism 
has extensively been investigated 
experimentally and theoretically from the viewpoint of fundamental few-body problems 
in nuclear physics and atomic/molecular 
physics~\cite{Breunlich89,Ponomarev90,Froelich92,Bogdanova1988,Nagamine98}.

An example of interesting points of studying the $dd\mu$ fusion is 
to examine the violation of the charge symmetry between reactions (1.1) and (1.2) in 
the $p$-wave component, since reactions (1.3) and (1.4) take place purely in the 
$p$-wave $d$-$d$ relative motion as mentioned above. 
For this purpose, the following two kinds of ratios have been studied,
%
\setcounter{equation}{4} 
\begin{eqnarray}
  &&  R_S = S(^3{\rm He}+n)/S(t+p),  \\
  &&   R_Y= Y(^3{\rm He}+n)/Y(t+p),
\end{eqnarray}
where $S$ is the $p$-wave contribution of the astrophysical $S(E)$-factor of 
the reaction (1.1) or (1.2) at the $d$-$d$ center of mass (c.m.) energy $E \to \!0$, 
whereas $Y$ is the yield of the $dd\mu$ fusion reaction (1.3) or (1.4).
$R_S=R_Y=1.0$ is expected in the purely charge symmetric case.
Note that the ratio $R_S$ is the same as that of the $p$-wave {\it cross sections} 
at $E \to 0$ (cf. Eq.~(\ref{eq:S-factor})).

Bogdanova {\it et al.} (1982) pointed out that the yield ratio $R_Y$ is equal to $R_S$ 
at $E \to 0$ under the factorization assumption of the $dd\mu$ fusion rate as in Eq.~(4) 
of Ref.~\cite{Bogdanova1982}, where a large asymmetry $R_S=1.46$ was cited from 
the observation by Adyasevich {\it et al.}~\cite{Adya1981} (1981) at $E \to 0$.
From the $dd\mu$ fusion experiment, Balin {\it et al.}~\cite{Balin1984} (1984) obtained 
$R_Y=1.39 \pm 0.04$. 
By the $R$-matrix calculation of the four-nucleon system, 
Hale~\cite{Hale1990} (1990) presented  $R_S=1.43$. 
In the new experiment by Balin {\it et al.}~\cite{Balin2011}  (2011), $R_Y=1.445 \,(11)$ 
was reported.

\begin{figure}
\setlength{\abovecaptionskip}{0.3cm}
\setlength{\belowcaptionskip}{-0.cm}
\centering
\includegraphics[width=0.48\textwidth]{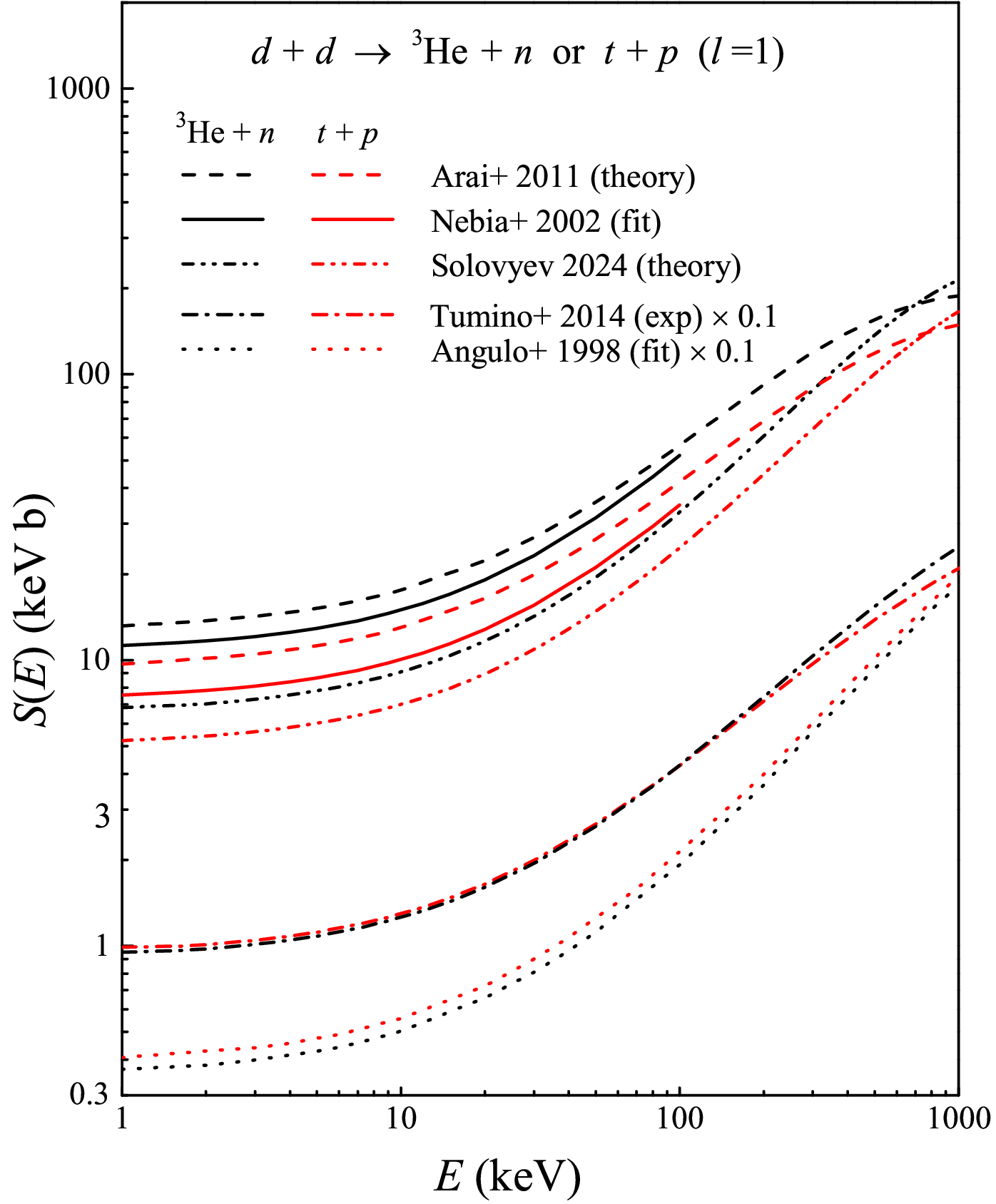}
\caption{
$p$-wave astrophysical $S(E)$ factors of reactions (1.1) and (1.2), reported by 
Angulo and Decouvemont~\cite{Angulo1998} (Angulo+), 
Nebia {\it et al.}~\cite{Nebia2002} (Nebia+),
Arai {\it et al.}~\cite{Arai2011} (Arai+),
Tumino {\it et al.}~\cite{Tumino2014} (Tumino+), 
and Solovyev~\cite{Solovyev2024} (Solovyev). 
Tumino+ and Angulo+ have been multiplied by 0.1 to avoid crowds of lines.
\mbox{Nebia+ is up to 100 keV.
No result reports error bar.
}
}
\label{fig:8line-sfactor}
\end{figure}

Up to now, there have appeared interesting experimental and theoretical studies on the 
$p$-wave astrophysical $S(E)$ factors of the reactions (1.1) and (1.2), in a broad range 
of the center-of-mass energy \mbox{$E \simeq$ 1 keV} to 1 MeV, by five 
groups~\cite{Angulo1998,Nebia2002,Arai2011,Tumino2014,Solovyev2024}.  
However, the results  are significantly different 
from each other as illustrated in Fig.~\ref{fig:8line-sfactor}, and have not been used yet 
in the study of the $dd\mu$ fusion.

\begin{figure}
\setlength{\abovecaptionskip}{0.cm}
\setlength{\belowcaptionskip}{-0.cm}
\centering
\includegraphics[width=0.38\textwidth]{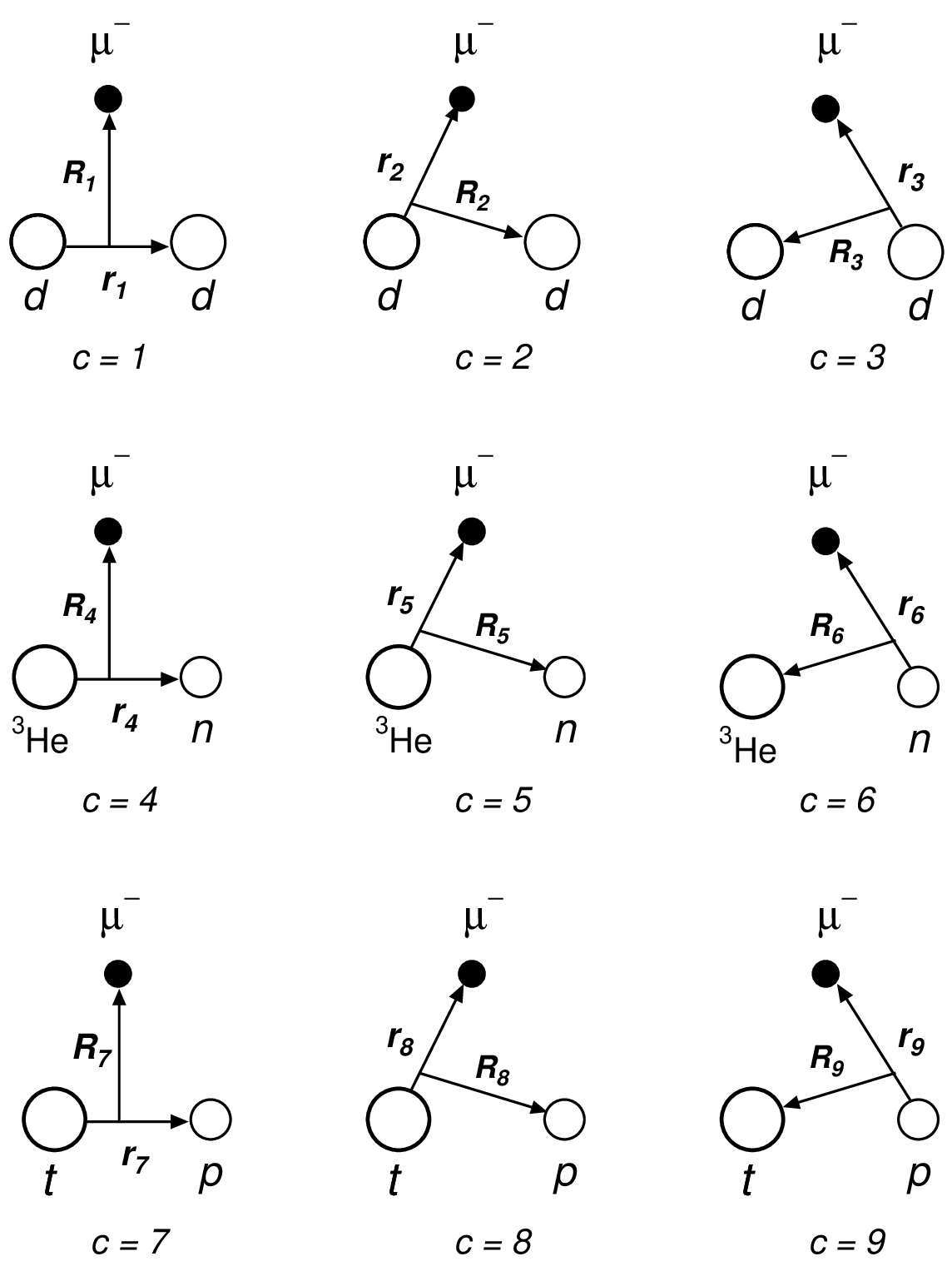}
\vspace*{0.5cm}
\caption{Nine Jacobi coordinates used in this work for the $dd\mu$, $^3{\rm He}n\mu$, 
and $tp\mu$ systems, referred to as channel $c=1$ to $c=9$, respectively.}
\label{fig:3body-jacobi}
\end{figure}

Thus, the purpose of the present work is that, analyzing the $p$-wave  $S(E)$ factors 
in Fig.~\ref{fig:8line-sfactor} for the first time, we comprehensively study  
the reaction processes in the $dd\mu$ fusion on the basis of the $T$-matrix 
method~\cite{Wu2024}, with the use of nine sets of the Jacobi coordinates (channels) 
in Fig.~\ref{fig:3body-jacobi}. 
Including the above-mentioned charge-symmetry violation,
we investigate the $dd\mu$ fusion rates, the $\mu$-$^3{\rm He}$ sticking probabilities, 
and the energy (momentum) spectra of the emitted muon.

%
 
We employ three kinds of methods to derive the fusion rate of the $dd\mu$ molecular state, 
and show that their results are consistent with each other. 
The present study proceeds along the following {\it steps} 1) to 4):

{\it Step} 1) \: 
Reproduce the $p$-wave $S(E)$ factors 
by employing the optical-potential model, which is successfully used for the $dt\mu$ 
fusion~\cite{Kamimura2023,Wu2024,Kamimura:1989AIP}. 
The so-obtained complex $d$-$d$ potential  is then used when calculating the \mbox{$J=1$} 
states of the $dd\mu$ molecule. The fusion rate of the molecular state 
is given by using the imaginary part of the complex eigenenergy.
This optical-potential method is referred to as \mbox{{\it method} i).}

{\it Step} 2) \: 
To calculate the reaction rates of (1.3) and (1.4), while
taking into their outgoing channels explicitly, 
we employ the tractable $T$-matrix model~\cite{Wu2024}. 
We determine the nonlocal coupling potential between 
the $d$-$d$ and $^3{\rm He}$-$n$ \,($t$-$p$) channels
so that using the $T$-matrix model can reproduce individually the five kinds of the 
$p$-wave $S(E)$ factors
in Fig.~\ref{fig:8line-sfactor}.

{\it Step} 3) \: 
Then, use of the so-obtained potential sets in the \mbox{$T$-matrix} model~\cite{Wu2024} 
for the reactions (1.3) and (1.4) can result in the reaction (fusion) rates 
that are consistent among the selected potential sets. 
This $T$-matrix model calculation performed on channels 5 and 8 (Fig.~\ref{fig:3body-jacobi}) 
of the outgoing waves is referred to as {\it method} ii), while that on channels 4 and 7 
as {\it method} iii).

{\it Step} 4) \: 
Derive the $^3{\rm He}$-$\mu$, $t$-$\mu$ and $p$-$\mu$
sticking probabilities using the absolute values of the reaction rates to all the outgoing 
channels of (1.3) and (1.4) obtained by {\it method} ii).
Furthermore, calculate the absolute strengths of the 
momentum and energy spectra of the emitted muon in (1.3) and (1.4) with \mbox{{\it method} iii)}. 
Muon spectrum reflects the nature of $dd\mu$ molecule wave function before the fusion reaction.

This paper is organized as follows:
In Sec.~\ref{secII}, using \mbox{{\it method} i),}
we calculate the $p$-wave $S(E)$ factors and the fusion rate of the $(dd\mu)_{J=1}$ molecule.
In Sec.~III, the coupling potential between  $d$-$d$ and $^3{\rm He}$-$n$ ($t$-$p$) 
channels is determined.
In Sec.~IV, employing {\it method} ii), we calculate 
the fusion rate of the $(dd\mu)_{J=1}$ state together with the reaction rates to the 
outgoing continuum and bound states of the reactions (1.3) and (1.4).
In Sec.~V, using these results, we derive the $^3{\rm He}$-$\mu$, $t$-$\mu$ and $p$-$\mu$ 
sticking probabilities.
In Sec.~VI, the fusion rates are calculated  using  {\it method} iii).
Spectra of the muons emitted is calculated in Sec.~VII. 
Charge-symmetry violation in the $p$-wave $d$-$d$ reaction and the $dd\mu$ fusion 
reaction is investigated in Sec.~VIII.
At last, a summary is presented in Sec.~IX.


\section{Fusion rate of \boldmath{\lowercase{$dd\mu$}} molecule (\lowercase{i}): \\
\vskip 0.05cm
Optical-potential model}\label{secII}

We firstly investigate the fusion reactions (1.1)-(1.4) using {\it method} i), 
as in Refs.~\cite{Kamimura2023,Wu2024,Kamimura:1989AIP}.
In all {\it methods} i) to iii), we select nuclear interactions in order to reproduce 
the $p$-wave  $S(E)$ factors in Fig.~\ref{fig:8line-sfactor}, \mbox{in which} 
Angulo and Descouvemont~\cite{Angulo1998} made the \mbox{$R$-matrix} parametrization 
fit to the observed data by Refs.~\cite{Schulte1972,Kraus87,Brown1990,Gleife1995}, 
Nebia {\it et al.}~\cite{Nebia2002} analyzed experimental 
studies~\cite{Kraus87,Brown1990,Gleife1995,Bosch1992,Theus1966} \mbox{using} the WKB 
approximation ($E \le$ 100 keV),
Arai {\it et al.}~\cite{Arai2011} performed {\it ab initio} four-nucleon calculation 
with a realistic $NN$ force AV8$'$~\cite{AV8}, Tumino {\it et al.}~\cite{Tumino2014}
obtained the experimental data using the Trojan Horse method, and 
Solovyev~\cite{Solovyev2024} employed a microscopic multichannel cluster model taking 
a semirealistic effective $NN$ force~\cite{Kaneko}.

\subsection{Parameters to reproduce 
\boldmath {$p\,$}-wave $S(E)$ factors}

The parameters of the nuclear $d$-$d$ potential are determined by reproducing 
the {\it summed} cross sections of reactions (1.1) and (1.2).
The total angular momentum and parity $I^\pi$ are $I^\pi=0^-, 1^-$ and $2^-$ 
with \mbox{$p$-wave} $(l=1)$ and spin $S=1$. 

We describe the $d$-$d$ scattering
wave function $\Phi_{dd,\,I M}^{\rm (opt)}(E, {\bf r})$
at the c.m. energy $E$ as (with obvious notations),
\begin{eqnarray}
\Phi_{dd,\,I M}^{\rm (opt)}(E, {\bf r})
= \phi_{dd, l}^{\rm (opt)}(E, r)  \left[ Y_l({\hat{\bf r}}) \, 
  \chi_{S}(dd) \right]_{I M} .  \quad \quad
\end{eqnarray}
Schr\"{o}dinger equation for $\phi_{dd, I M}^{\rm (opt)}(E, {\bf r})$  
is,
\begin{eqnarray}
\label{eq:opt-Sch-2body}
 &&   ( H_{dd} - E)\, \phi_{dd, l}^{\rm (opt)}(E,r) \,Y_{lm}({\hat{\bf r}}) =0, \\
\label{eq:opt-Hamil}
 &&  H_{dd} = T_{{\bf r}}
             + V_{dd}^{{\rm (N})}(r) + iW_{dd}^{{\rm (N})}(r)
          + V_{dd}^{({\rm Coul})}(r),  \quad \quad \quad \quad 
\end{eqnarray}
where we assume the following $d$-$d$ potential for $l=1$ and $S=1$
(independent of $I$), 
with usual notations,
\begin{eqnarray}
\label{eq:opt-real}
    V_{dd}^{{\rm (N})}(r)&\! =\! &V_0/\{1+e^{(r-R_0)/a}\} , \\
\label{eq:opt-imag}
   W_{dd}^{{\rm (N})}(r) &\! =\! & W_0/\{1+e^{(r-R_{\rm I})/a_{\rm I}} \} , \\
\label{eq:opt-coul}
 V_{dd}^{({\rm Coul})}(r) &\! =\! & \begin{cases}
           (e^2/(2R_{\rm c}))(3-r^2/R_{\rm c}^2)\,,
       &  \text{$ \; r < R_{\rm c}$},  \\
          e^2/r\,,  &  \text{$ \; r \geq  R_{\rm c}$} \quad \quad \quad \quad 
\end{cases}
\end{eqnarray}
while taking a fixed charge radius $R_{\rm c}=3.0$ fm.

It is to be noted that, in the energy regions of 
Fig.~\ref{fig:8line-sfactor}, only the two outgoing channels of reactions (1.1) 
and (1.2) are open except for the incoming channel.
Therefore, the absorption cross section for $l=1$ 
is nothing but the \mbox{$p$-wave} one, $\sigma(E)$,
which is represented as
\begin{eqnarray}
\sigma(E)= C_{l,S} \frac{\pi}{k^2} (1-|S_l|^2), \quad (l=1)
\end{eqnarray}
with $S_l$ the $S$-matrix, and
\begin{eqnarray}
C_{l,S}=\frac{(2l+1)(2S+1)(1+\delta)}{(2I_d+1)(2I_d+1)}=1,  \;\; (S=I_d=1),  \;\;  \;\; 
\end{eqnarray}
where $\delta=1$ for the identical colliding particles, and 
$k$ being the wave number of the $d$-$d$ relative motion.
The corresponding \mbox{$S(E)$ factor}  is 
\mbox{derived} from the cross section as
\begin{eqnarray}
\sigma(E)=S(E)\, e^{-2\pi \eta(E)}/E,
\label{eq:S-factor}
\end{eqnarray}
where $\eta(E)$ denotes the Sommerfeld parameter.

\begin{figure}
\centering
\includegraphics[width=0.46\textwidth]{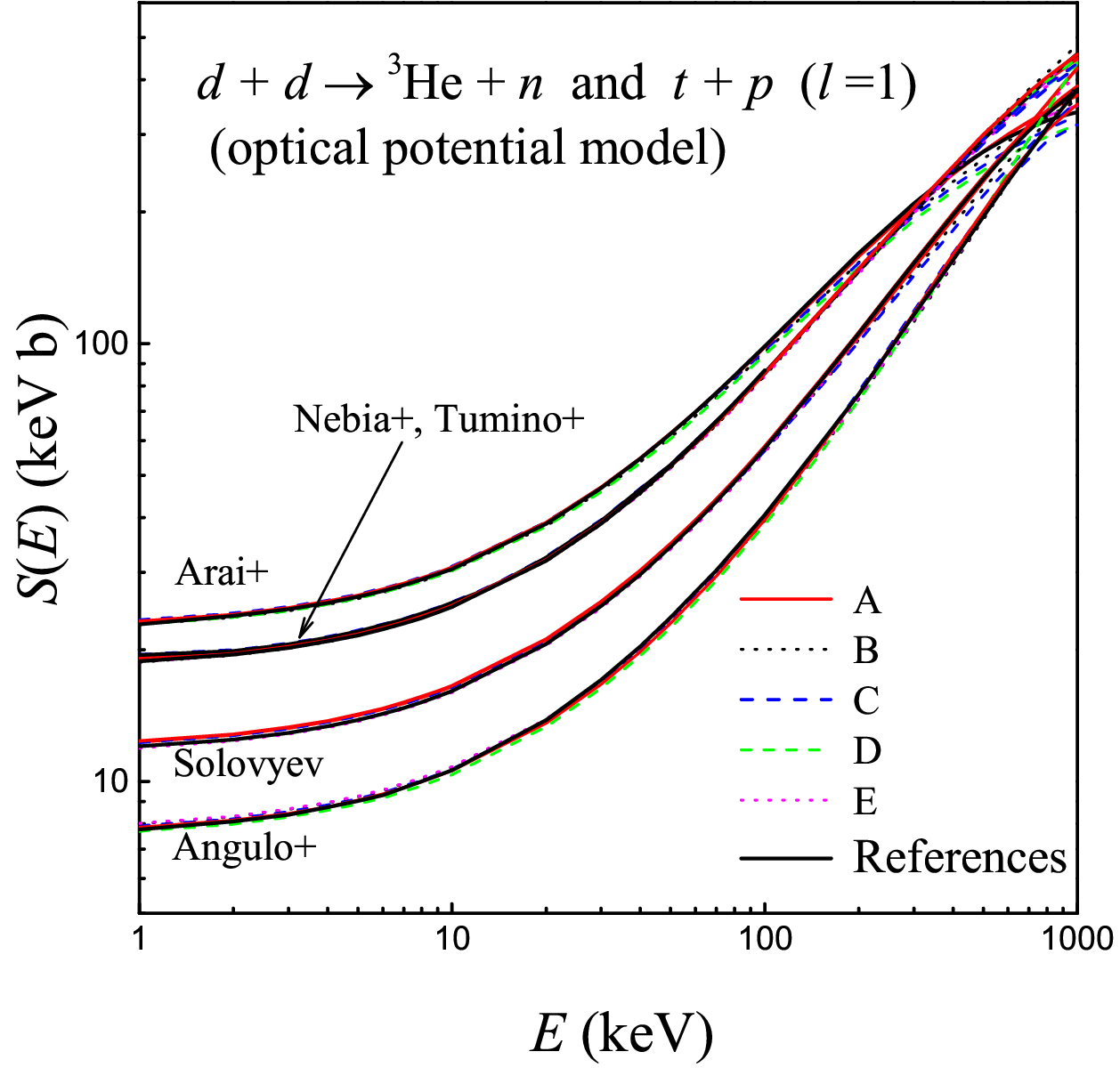}
\caption{
$p$-wave $S$-factor $S(E)$, with black lines given as sum of the two $S(E)$ of 
reactions (1.1) and (1.2) for each reference in Fig.~\ref{fig:8line-sfactor}.
Each black line is well reproduced by the five sets 
$d$-$d$ optical-potentials listed in Table~\ref{tab:vdd}. 
}
\label{fig:opm-S}
\end{figure}

In order to analyze the $S(E)$ factors in Fig.~\ref{fig:8line-sfactor}, we sum 
the two lines for  $^3{\rm He}+n$ and $t+p$ with respect to each reference and 
illustrate as black solid lines in Fig.~\ref{fig:opm-S}.
Since the optical-potential is phenomenological, we select 
five quite  different sets A to E, all well reproducing the individual black line 
in Fig.~\ref{fig:opm-S}. 
The potential parameters are listed in Table \ref{tab:vdd} for the cases of 
Tumino+ and Nebia+, but  $V_0$ and $W_0$ 
for Angulo+, Arai+, and Solovyev are not listed for simplicity. 
  
\begin{table}
\centering
\caption{
Five sets (A to E) of the $d$-$d$ optical-potential parameters we used to fit 
the $p$-wave $S(E)$ factor of Tumino+ 2014, and then others. The numbers in 
the parentheses are for Nebia+ 2002.
$V_0$ and $W_0$ for Angulo+ 1998, Arai+ 2011 and Solovyev 2024 
are not written to prevent complexity. 
$R_{\rm c}=3.0$ fm for all.
}
\begin{tabular}{ccccccc}
       \hline
       \hline
       \noalign{\vskip 0.1 true cm}
        Pot.   & $V_0$ & $W_0$ &$R_0$ & $a$&  $R_I$ & $a_I$ \\
        set  & (MeV) & (MeV) & (fm) & (fm) & (fm) & (fm) \\
          \noalign{\vskip 0.1 true cm}
        \hline
        \noalign{\vskip 0.1 true cm}
        A & $-12.60 \: (-13.10) $ &$\;\;-1.04 \: (-1.02) $ & 
                \;\;6.0 &\;\;0.9 &\;\;3.0 &\;\;0.9 \\
        B & $-14.40 \: (-14.10) $ &$\;\;-1.10 \: (-1.08) $ & 
                \;\;6.0 &\;\;0.3 &\;\;6.0 &\;\;0.3 \\
        C & $-22.00 \: (-22.10) $ &$\;\;-0.70 \: (-0.70) $ & 
                \;\;4.0 &\;\;1.0 &\;\;5.0 &\;\;1.0 \\ 
        D & $-29.80 \: (-29.90) $ &$\;\;-1.90 \: (-1.96) $ & 
                \;\;3.0 &\;\;0.3 &\;\;5.0 &\;\;0.3 \\
        E & $-36.00 \: (-36.90) $ &$\;\;-1.50 \: (-1.42) $ & 
                \;\;6.0 &\;\;0.5 &\;\;5.0 &\;\;0.5 \\
        \noalign{\vskip 0.1 true cm}
        \hline
        \hline
    \end{tabular}
    \label{tab:vdd}
\end{table}
%

\subsection{Fusion rate of the \boldmath{$dd\mu$} molecule}

We then solve the three-body Schr\"{o}dinger equation 
for the $(dd\mu^-)_{J v}$ states including the
nuclear optical potential obtained above.  In this model,
the fusion rate of the reactions (1.3) and (1.4), say $\lambda_{J v}$, 
are derived from the imaginary part of the complex eigenenergy $E_{J v}$.

We perform a non-adiabatic three-body calculation 
of  the excited states with $J=1$, 
using the Gaussian Expansion Method \mbox{(GEM) for
few-body systems~\cite{Kamimura1988zz,Kameyama89,Hiyama03}.}  
The Schr\"{o}dinger equation for the wave functions $\Phi^{\rm (opt)}_{J M}(dd\mu)$ 
and eigenenergies $E_{J}$ are given by 
\begin{eqnarray}
\label{eq:Sch-dtmu-bound}
   && ( H_{dd\mu} - E_{J})\, \Phi^{\rm (opt)}_{JM}(dd\mu) =0,  \\
\label{eq:Hamil-ddmu-bound}
   &&  H_{dd\mu}=  T_{{\bf r}_c} +  T_{{\bf R}_c}
          + V^{({\rm C})}(r_2) + V^{({\rm C})}(r_3)        \nonumber \\
    && \qquad         + V^{({\rm N})}_{dd}(r_1) + iW^{({\rm N})}_{dd}(r_1)
         + V^{({\rm C})}_{dd}(r_1)\,.
\end{eqnarray}
$\Phi^{\rm (opt)}_{J M}(dd\mu)$ is constructed as the sum of
amplitudes of the three rearrangement channels c=1, 2, and 3
in Fig.~\ref{fig:3body-jacobi},
\begin{eqnarray}
\!\!
{\Phi}^{\rm (opt)}_{J M}(d d \mu) &\!\!\!\!=\!\!\!\!& 
 \Phi_{JM}^{(1)}\!\left(\mathbf{r}_1, \mathbf{R}_1\right) \nonumber \\
&&\!\!\!\!\!\!\!\!+
\left[\Phi_{JM}^{(2)}\!\left(\mathbf{r}_2, \mathbf{R}_2\right) +
\Phi_{JM}^{(3)}\!\left(\mathbf{r}_3, \mathbf{R}_3\right)\right] .  
\label{eq:gemwf}
\end{eqnarray}
$\Phi_{JM}^{(2)}$ and $\Phi_{JM}^{(3)}$ are symmetrized between two deuterons.
The amplitude of each channel $c$ is expanded in terms of Gaussian basis functions of 
the Jacobian coordinates 
\mbox{${\bf r}_c$ and ${\bf R}_c \:(c=1-3)$,}
\begin{eqnarray}
\Phi_{JM}^{(c)}({\bf r}_c, {\bf R}_c) =
 \!\!\!\! \sum_{n_s l_c, N_c L_c} \!\!\!\!A_{n_c l_c, N_c L_c}^{(c)} 
\left[ \phi_{n_c l_c}({\bf r}_c) \psi_{N_c L_c}
({\bf R}_c) \right]_{JM},  \nonumber \!\!\!\!\!\! \!\! \\  
\end{eqnarray}
where the basis functions and their amplitudes are symmetric between the channels $c=2$ and $c=3$.
The basis functions are given by 
\begin{equation}
\begin{aligned}
& \phi_{n l m}(\mathbf{r})=\phi_{n l}(r) Y_{l m}(\hat{\mathbf{r}}), \\
& \phi_{n l}(r)=N_{n l} r^l e^{-\nu_n r^2}, \quad\left(n=1-n_{\max }\right), \\
& \psi_{N L M}(\mathbf{R})=\psi_{N L}(R) Y_{L M}(\widehat{\mathbf{R}}), \\
& \psi_{N L}(R)=N_{N L} R^L e^{-\lambda_N R^2}, \quad\left(N=1-N_{\max }\right),
\end{aligned}
\label{eq:gaussian-basis}
\end{equation}
with  normalization constants $N_{n l}$ and $N_{N L}$.
Gaussian range parameters $\nu_n$ and $\lambda_n$ 
are postulated to lie in geometric progression,
\begin{equation}
\begin{aligned}
& \nu_n=1 / r_n^2, \quad r_n=r_1 a^{n-1},\left(n=1-n_{\max }\right), \\
& \lambda_N=1 / R_N^2, \quad R_N=R_1 A^{N-1},\left(N=1-N_{\max }\right) .
\label{eq:gaussian-range}
\end{aligned}
\end{equation}
The eigenenergy and  wave function are obtained
using the Rayleigh-Ritz variational method.

As the eigenenergy $E_{Jv}$ is a complex number, we write
$E_{Jv}=E_{Jv}^{(\rm real)}+i E_{Jv}^{(\rm imag)}$ and
introduce $\varepsilon_{Jv}=E_{Jv}^{(\rm real)}-E_{\rm th}$, with $E_{\rm th}$ being 
the $(d\mu)_{1s}+d$ threshold energy.
The diagonalization in the cases of $l_{\rm max}=4$ ($l_{\rm max}=2$)
yields  $\varepsilon_{10}=-226.679\, (-226.665$) eV and 
$\varepsilon_{11}=-1.974 \, (-1.961)$ eV.
Contribution from the nuclear interaction is $-1.44 \times 10^{-6}$ eV in the real 
part and $-1.39 \times 10^{-7}$ eV 
in the imaginary part.
According to Ref.~\cite{Kamimura2023}, the digits below 1 eV in the real part 
did not affect the fusion reaction calculation.
Thus, we employ $l_{\rm max}=2$ in this work. The input Gaussian basis is 
shown in Table~\ref{tab:gaussian-para}.
We take seven lines of the Gaussian basis parameters
where the third line for $c=1$ is effective to the $d$-$d$ nuclear interaction. 

Here, we note that
the GEM calculation is transparent in the sense that
{\it all} the nonlinear variational parameters can explicitly be
reported in a small table such as Table~II.
Since the computation time required for calculating the Hamiltonian 
matrix elements with the Gaussian basis set is very short, 
we can take an appropriately large number (even more than enough) 
of basis functions.
Use of this very wide function space constructed on all the 
three Jacobi coordinates facilitates the 
ease of optimization of the Gaussian ranges using {\it round} numbers
such as those presented in Table~II; cf. other advantages of the GEM calculations  
shown in the \mbox{review paper~\cite{Hiyama03}}.

\begin{table}
\centering
\caption{All the nonlinear variational parameters of the  Gaussian basis functions, 
with $J=1$ in Eqs.~(\ref{eq:gaussian-basis})--(\ref{eq:gaussian-range}).
$r_1$($R_1$) and $r_{\rm max}$($R_{\rm max}$)
are in units of $a_{\mu}=\hbar^2/m_\mu e^2= 255.9$ fm.
2,600 basis functions in total.}
\begin{tabular}{p{1.0cm}<{\centering}p{0.5cm}<{\centering}p{0.5cm}
<{\centering}p{1.0cm}<{\centering}p{1.0cm}<{\centering}p{0.5cm}
<{\centering}p{0.6cm}<{\centering}p{1.0cm}<{\centering}p{0.8cm}<{\centering}}
\noalign{\vskip 0.1 true cm}
\hline \hline
\noalign{\vskip 0.1 true cm}$\mathrm{c}$ & $l_c$ & $n_{\max }$ & $\begin{array}{c}r_1 \\
{[a_{\mu}]}\end{array}$ & $\begin{array}{c}r_{n_{\max }} \\
{[a_{\mu}]}\end{array}$ & $L_c$ & $N_{\max }$ & $\begin{array}{c}R_1 \\
{[a_{\mu}]}\end{array}$ & $\begin{array}{c}R_{N_{\max }} \\
{[a_{\mu}]}\end{array}$ \\
\noalign{\vskip 0.1 true cm}
\hline
\noalign{\vskip 0.1 true cm}
1 & 1 & 25 & 0.05 & 10 & 0 & 15 & 0.1 & 15 \\
1 & 1 & 15 & 0.05 & 10 & 2 & 15 & 0.1 & 15 \\
1 & 1 & 25 & 0.001 & 0.05 & 0 & 15 & 0.1 & 15 \\
2, 3 & 0 & 20 & 0.02 & 15 & 1 & 15 & 0.1 & 25 \\
2, 3 & 1 & 15 & 0.02 & 10 & 0 & 15 & 0.1 & 25 \\
2, 3 & 2 & 15 & 0.02 & 10 & 1 & 15 & 0.1 & 25 \\
2, 3 & 1 & 15 & 0.02 & 10 & 2 & 15 & 0.1 & 25 \\
\noalign{\vskip 0.1 true cm}
\hline
\hline
\end{tabular}
\label{tab:gaussian-para}
\end{table}

Nuclear fusion occurs nearly exclusively from the 
\mbox{$J\!=\!v\!=1$} states, as pointed out by Balin {\it et al.}~\cite{Balin2011};
in the symmetric $dd\mu$ molecule, the $\Delta J=1$ transitions are forbidden, 
apart from small relativistic effects. The calculated \mbox{$\Delta J=0$} 
\mbox{deexcitation} rate, $\Gamma_{\rm dex}$, 
from the $J=\!v\!=1$ to the $v=0$ states is 
$\Gamma_{\rm dex}=0.2 \times 10^8 {\rm s}^{-1}$~\cite{Bakalov1988}, 
which is rather low compared with the theoretical fusion rates 
$\lambda_{11}=0.44 \times 10^9 {\rm s}^{-1}$  and
$\lambda_{10}=1.5 \times 10^9 {\rm s}^{-1}$ 
by Bogdanova {\it et al.}~\cite{Bogdanova1982}.
Therefore, in the following, we treat the results for the fusion from the 
$J=v=1$ state more importantly than those for the $J=1, v=0$ state.

The latest observed value of the effective fusion rate,
${\tilde \lambda}_{\rm f}$, was given 
by Balin {\it et al.}~\cite{Balin2011}
as ${\tilde \lambda}_{\rm f}=(3.81 \pm 0.15) \times 10^8 {\rm s}^{-1}$ 
for which ${\tilde \lambda}_{\rm f}$ is defined by
\begin{equation}
 \widetilde{\lambda}_{\rm f} = \lambda_{11} + \Gamma_{\rm dex} .
\label{eq:effective-fusion}
\end{equation}
Other effective fusion rates given by Petitjean {\it al.}~\cite{Petitjean1999}
and by Voropaev {\it et al.}~\cite{Voropaev2001} are listed
in \mbox{Table~\ref{tab:lambda-ref}}, together with the calculated literature 
values by Refs.~\cite{Bogdanova1982,Hale1990,Alexander1990} in which 
$\lambda_{11}$  were calculated by using the $p$-wave $S(E)$ factor
at $E \to 0$ obtained by Ref.~\cite{Adya1981} (1981) 
on the basis of the factorization method for the fusion 
\mbox{reactions~\cite{Jackson1957,Bogdanova1982}.} 

In the present optical-potential model,
the fusion rate, $\lambda^{\rm (opt)}_{Jv}$, of reactions (1.3) and (1.4) is derived 
by the inverse of the lifetime of the molecular state,
\begin{equation}
\lambda^{\rm (opt)}_{Jv}=-2E_{Jv}^{({\rm imag})}/\hbar, 
\end{equation}
and is listed in Table~\ref{tab:lambda-opm} for the five cases of $S(E)$ factors, 
averaged over the optical-potential sets A to E.
We see that, for each $S(E)$, quite different potential sets 
generate almost the same fusion rates with very small deviation. This clearly
shows the validity of our optical-potential model for the present subject. 

\begin{table}
\centering
\caption{
Literature results for the fusion rate $\lambda_{11}$ 
and the effetive fusion rates $\widetilde{\lambda}_{\rm f}$
by calculations (the upper three) and by experiments (the lower three).
The numbers with \mbox{superscript(*)} are estimated using the definition 
Eq.~(\ref{eq:effective-fusion}). All the rates are in units of $10^8{\rm s}^{-1}$.
}
\begin{tabular}{lclcl}
       \noalign{\vskip 0.1 true cm}
       \hline
       \hline
\noalign{\vskip 0.1 true cm}
     & \qquad \qquad & $\lambda_{11}\quad$   
        &\qquad \qquad  & $\quad \widetilde{\lambda}_{\rm f} \qquad$  \\
\noalign{\vskip 0.1 true cm}
\hline
\noalign{\vskip 0.1 true cm}
(Cal.)
  & & & & \\
\noalign{\vskip 0.05 true cm}
 Bogdanova {\it et al.} (1986)&  &   $4.4$ &  &  $4.6^*$ \\
\noalign{\vskip 0.1 true cm}
 Hale  (1990/1991) & &    $3.6^*$ & & $3.8$ \\
\noalign{\vskip 0.1 true cm}
 Alexander {\it et al.} (1991)  & &   $ 3.8 $ & & $4.0^*$ \\
\noalign{\vskip 0.1 true cm}
(Exp.)  & & & & \\
\noalign{\vskip 0.05 true cm}
 Petitjean {\it et al.} (1999)&  &  $3.3^*$  &  &   $3.5$ \\
\noalign{\vskip 0.1 true cm}
 Voropaev {\it et al.} (2001) & & $3.9^*$ & &  $4.07(20)$  \\
\noalign{\vskip 0.1 true cm}
 Balin {\it et al.} (2011)  & &  $ 3.6^* $ & &  $3.81(15)$ \\
\noalign{\vskip 0.1 true cm}
\noalign{\vskip 0.05 true cm}
\hline
\hline
\end{tabular}
\label{tab:lambda-ref}
\end{table}

\begin{table}
\centering
\caption{
Fusion rate $\lambda^{\rm (opt)}_{Jv}$ of the $(dd\mu)_{Jv}$ states 
(\mbox{$J=1,$} $v=1, 0)$ for the reactions (1.3) and (1.4), calculated and 
averaged over the five \mbox{$d$-$d$} optical-potentials sets (Table \ref{tab:vdd}).
Same meaning for the numbers with \mbox{superscript(*)} as in 
Table \ref{tab:lambda-ref}. All the rates are in units of $10^8{\rm s}^{-1}$.
}
\begin{tabular}{cccc}
       \noalign{\vskip 0.1 true cm}
       \hline
       \hline
\noalign{\vskip 0.1 true cm}
 $p$-wave $S(E)$ factor     & $\qquad \lambda^{\rm (opt)}_{11} \quad$ 
          & $\qquad \widetilde{\lambda}^{\rm (opt)}_{\rm f} \qquad$ 
     & $\quad\lambda^{\rm (opt)}_{10}\quad$   \\
\noalign{\vskip 0.1 true cm}
\hline
\noalign{\vskip 0.1 true cm}
 \;\;   Angulo+ 1998 \;\;  &  $\;\;1.75(5)$ \footnote{
The numbers, for example, 1.75(5), means that the deviations from 
the average 1.75 with respect to the five potential sets is within a range 
of $\pm 0.05$. The same apply in such expressions for numerical results of 
the present work.} 
&  $2.0^* \;\;\;$  & $\;\;\,\;\,5.5(2)$  \\
\noalign{\vskip 0.1 true cm}
    Nebia+ 2002       &   $4.16(4)$ &  $4.4^* \;\;\;$    & $\,\;\,12.9(1)$  \\
\noalign{\vskip 0.1 true cm}
    Arai+ 2011       &   $5.05(5)$ &  $5.3^* \;\;\;$    & $\,\;\,15.7(1)$  \\
\noalign{\vskip 0.1 true cm}
    Tumino+ 2014     &   $4.21(3)$ &  $4.4^* \;\;\;$  & $\,\;\,13.1(1)$  \\
\noalign{\vskip 0.1 true cm}
    Solovyev 2024     &    $2.69(5)$ &  $2.9^* \;\;\;$  & $\;\;\,\;\,8.4(2)$  \\
\noalign{\vskip 0.1 true cm}
\hline
\hline
\end{tabular}
\label{tab:lambda-opm}
\end{table}

The difference in the fusion rates $\lambda^{\rm (opt)}_{J=1,v}$
among the five cases  in Table \ref{tab:lambda-opm}
reflects the difference in the \mbox{$p$-wave}$p$-wave  $S(E)$ factor in 
Fig.~\ref{fig:opm-S} for the reactions (1.1) and (1.2).
The fusion rates distribute in a wide range in Table IV although including the 
observed values and calculated literature values in Table III.
More discussions on the fusion rates will be made in Secs. IV and VI employing
our $T$-matrix model.

\vskip 0.7cm
\section{ {\boldmath $T$}-matix model for
\mbox{\boldmath {$\lowercase{d + d} \to \,\!^3\mbox{H\lowercase{e}} + \lowercase{n}
 $}} \, \\ \lowercase{and} \,
  \mbox{\boldmath {$ \lowercase{d + d} \to \lowercase{t} + \lowercase{p} $}    } \:
}

In Sec.~III of Ref.~\cite{Wu2024}, we analyzed the $S(E)$-factor 
of the $d + t \to\!\!~^4{\rm He} + n$ reaction for $E=1$ to 300 keV
using the tractable $T$-matrix model.
In this section, we perform a similar analysis of the reactions (1.1) and (1.2), 
in which the incoming wave has $l=1$ and $S=1$.
We determine the potential parameter sets via reproducing the five cases of 
\mbox{$p$-wave} \mbox{$S(E)$ factors}~\cite{Angulo1998,Nebia2002,Arai2011,
Tumino2014,Solovyev2024}, as illustrated in Fig.~\ref{fig:8line-sfactor}. 
Here, the $d$-$d$, $^3{\rm He}$-$n$ and $t$-$p$ relative coordinates are 
referred to as ${\bf r}_1$, ${\bf r}_4$, and ${\bf r}_7$ 
(Fig.~\ref{fig:3body-jacobi}), respectively, similarly to the three-body case.

Referring to the $T$-matrix model used in the \mbox{$d+t \to \alpha + n$} 
reaction (Sec.~III of Ref.~\cite{Wu2024}), we describe the cross sections 
$\sigma_{\!dd\to^3{\rm He}n}(E)$ and   $\sigma_{\!dd\to tp}(E)$ as follows, 
with the notations corresponding to those in Eqs.~(3.7) and (3.8) of 
Ref.~\cite{Wu2024},
%
\begin{eqnarray}
\label{eq:sigma-Tmat-CC-2body-hn}
\!\!\!\!\!\!\!\! &&  \sigma_{\!dd\to^3{\rm He}n}(E)   \nonumber \\
\!\!\!\!\!\!\!\!   && = \! \frac{v_{r_4}}{v_{r_1}}
      \left( \frac{\mu_{r_4}}{2 \pi \hbar^2} \right)^2
    \!\!\sum_{m_{^3{\rm He}} m_n} \!
\int | \, T^{(^3{\rm He}n)}_{m_{^3{\rm He}} m_n}({\bf k}_4)\, |^2\, 
{\rm d}{\bf \hat{k}_4},  \;    \qquad \\  
\!\!\!\!\!\!\!\! &&T^{(^3{\rm He}n)}_{m_{^3{\rm He}} m_n}({\bf k}_4)    \nonumber \\
\!\!\!\!\!\!\!\!  &&= 
\langle \,e^{  i {\bf k}_4 \cdot {\bf r}_4 }\,
       \chi^{(^3{\rm He})}_{{\frac{1}{2}} m_{^3{\rm He}}} \chi^{(n)}_{{\frac{1}{2}} m_n}
   |\,  {\cal V}^{({\rm cp})}_{\!^3{\rm He}n,\, dd} \,| \, 
   \Phi_{dd, IM}^{\rm (opt)}(E,{\bf r}_1) \, \rangle .    \qquad \;\;
\end{eqnarray}
%
and
\begin{eqnarray}
\!\!\!\!\!\!\!\! &&  \sigma_{\!dd\to tp}(E)  \nonumber \\
\!\!\!\!\!\!\!\! &&  =\!  \frac{v_{r_{\,7}}}{v_{r_1}}
      \left( \frac{\mu_{r_1}}{2 \pi \hbar^2} \right)^2
\sum_{m_t m_p} \int |\, T^{(tp)}_{m_t m_p}({\bf k}_7)\, |^2\, 
{\rm d}{\bf \hat{k}_7},\qquad   \\
\!\!\!\!\!\!\!\! && T^{(tp)}_{m_t m_p}({\bf k}_7)   \nonumber \\
\!\!\!\!\!\!\!\!  &&   =  
\langle\,e^{  i {\bf k}_7 \cdot {\bf r}_{\,7} }\,
       \chi^{(t)}_{\frac{1}{2} m_t} \chi^{(p)}_{\frac{1}{2} m_p} \,
    |\,  {\cal V}^{({\rm cp})}_{\!tp,\, dd} \,| \, 
    \Phi_{dd, IM}^{\rm (opt)}(E,{\bf r}_1) \, \rangle . \;\; \quad \quad   
\label{eq:sigma-Tmat-CC-2body-dt}
\end{eqnarray}
The $S$-factors $S_{\!dd\to^3{\rm He}n}(E)$ and 
$S_{\!dd\to tp}(E)$ are derived from the corresponding cross 
sections using Eq.~(\ref{eq:S-factor}).

Here, we note that, in the initial {\it ket} vector of Eqs.~(3.2) and (3.4), the exact 
solution  of the CC Eqs.~(1.1) and (1.2) is approximated by 
$\Phi_{dd, IM}^{\rm (opt)}(E,{\bf r}_1)$ of (2.1),
in which the effect of the outgoing channels is reflected through the imaginary potential 
$W^{\rm (N)}_{dd}$ to a considerable extent.

In Eq.(3.2), ${\cal V}^{({\rm cp})}_{\!^3{\rm He} n,\, dd}$
is a nonlocal coupling potential between the $d$-$d$ and $^3{\rm He}$-$n$ channels
with $l=1$ as  
\begin{eqnarray}
{\cal V}^{({\rm cp})}_{\!^3{\rm He} n,\, dd}
 =\int {\rm d}{\bf r}_1 V^{({\rm cp})}_{\!^3{\rm He} n,\, dd}({\bf r}_4, {\bf r}_1),
\end{eqnarray}
and similarly for ${\cal V}^{({\rm cp})}_{\!tp,\, dd}$ in Eq.~(3.4) as
\begin{eqnarray}
  {\cal V}^{({\rm cp})}_{tp,\, dd}
 =\int {\rm d}{\bf r}_1 V^{({\rm cp})}_{\!tp,\, dd}({\bf r}_7, {\bf r}_1).
\end{eqnarray}
In Ref.~\cite{Wu2024}, for the study of the $d+t \to \alpha + n$ reaction, we assumed 
the tensor-form separable-nonlocal coupling potential.  
However, in the case of the present \mbox{$d+d$} reaction, 
we assume the following spin-independent separable-nonlocal potential with projecting 
$l=1$ state,
\begin{eqnarray}
  V^{({\rm cp})}_{^3{\rm He} n,\, dd}({\bf r}_4, {\bf r}_1)
&\!\! =\!\!  & v_{^3{\rm He} n,\, dd}^{({\rm cp})}\,   
          e^{- \mu_4 \,r_4^2 - \mu_1 r_1^2} \, \,   \nonumber \\
      &\times&  r_4 r_1\,  \Big[ Y_l({\hat {\bf r}}_4)\,Y_l({\hat {\bf r}}_1)\Big]_0,  \\
 V^{({\rm cp})}_{t p,\, dd}({\bf r}_7, {\bf r}_1)
&\!\!  =\!\!  & v_{t p,\, dd}^{({\rm cp})} \,  e^{- \mu_7 \,r_7^2 - \mu_1 r_1^2}
                     \nonumber \\
     &  \! \times \!& \,r_7 r_1 \, 
         \Big[ Y_l({\hat {\bf r}}_7)\,Y_l({\hat {\bf r}}_1)\Big]_0 .  
\label{eq:tensor-2}
\end{eqnarray}

The $p$-wave cross sections  $\sigma_{\!dd\to^3{\rm He}n}(E)$ and   
$\sigma_{\!dd\to^3{\rm He}n}(E)$ can be explicitly written as,
\begin{eqnarray}
 \sigma_{\!dd\to^3{\rm He}n}(E) & \!\! =\!\!& \frac{v_{r_4}}{v_{r_1}}
      \left( \frac{\mu_{r_4}}{2 \pi \hbar^2} \right)^{\!2}
      \left| \,v_{^3\!{\rm He} n,\, dd}^{({\rm cp})} 
      \, S_1^{\rm (cp)} F_1\, J_1 \,\right|^2 \! ,        \; \qquad \\
 \sigma_{\!dd\to tp}(E) &\!\!=\!\!& \frac{v_{r_{\,7}}}{v_{r_1}}
      \left( \frac{\mu_{r_{\,7}}}{2 \pi \hbar^2} \right)^{\!2}
      \left| \,v_{t p,\, dd}^{({\rm cp})}
      \,S_1^{\rm (cp)} F_1\, {\widetilde J}_1 \,\right|^2,  \qquad 
\label{eq:sigma-answer}
\end{eqnarray}
with
\begin{eqnarray}
 F_1 &\!\!=\!\!&
 \!\int_0^{\,\infty} \!\! \phi_{dd, 1}^{\rm (opt)}(E, r)\, r_1 \,
e^{-\mu_1 r_1^2} \,r_1^2 {\rm d}r_1, \\
J_1 &\!\!=\!\!& \frac{4\pi}{\sqrt{3}}    \int \!\!  r_4 \,j_1(k_4 r_4) \,
e^{-\mu_4 r_4^2}\, r_4^2 \,{\rm d} r_4,  \nonumber \\
&\!\!=\!\!& \frac{-1}{2\sqrt{3}} \left( \frac{\pi}{\mu_4} \right)^{\frac{3}{2}}\!
 \frac{k_4}{\mu_4}  \, e^{-\frac{\mu_4 k_4^2}{4}}\!, 
\end{eqnarray}
and similarly for ${\widetilde J}_1$ with changing the suffix 4 to 7.
Since Eqs.~(3.9) and (3.10) are independent of the total angular momentum $I M_I$ 
(with $l=S=1$) due to the spin-independent coupling potentials (3.5) and (3.6), 
it is not necessary to take the average over $I$ in the R.H.S. of Eqs.~(3.1) and (3.3). 
\mbox{In Eq. (3.11),} $\phi_{dt, 1}^{\rm (opt)}(E, r_1)$  is normalized asymptotically as
\begin{eqnarray}
\!\!\!\!\!\!\!\!\! \phi_{dt, 1}^{\rm (opt)}(E, r_1)
 \stackrel{r_1 \to \infty}{\longrightarrow} e^{i \sigma_1}
\frac{F_1(k,r_1)}{kr_1}   +  \mbox{(outgoing w.f.)}, \quad  \;\;
\end{eqnarray}
with the $p$-wave Coulomb regular function $F_1(k,r)$ and phase shift $\sigma_1$. 
$j_1(k_4 r_4)$ is the spherical Bessel function of order 1.

In Eqs.~(3.9) and (3.10), the  spin factor $S_1^{({\rm cp)}}$ is written 
\mbox{independently} of $M_S$ of the spin $S=1$ as
\begin{eqnarray}
\!\!\!\!\!\!\!  S^{{\rm (cp)}}_1 &=&
    \langle \,[ \chi^{(t)}_{\frac{1}{2}} \chi^{(p)}_{\frac{1}{2}}]_{1 M_S} 
         \,|\, [ \chi^{(d)}_{1} \chi^{(d)}_{1} ]_{1 M_S}   \rangle \nonumber   \\
   &=&  \langle \,[ \chi^{(^3{\rm He})}_{\frac{1}{2}} \chi^{(n)}_{\frac{1}{2}}]_{1 M_S} 
         \,|\, [ \chi^{(d)}_{1} \chi^{(d)}_{1} ]_{1 M_S}   \rangle,
\label{eq:spin-fac}
\end{eqnarray}
where we assume that the spin structure of the $^3{\rm He}$ and $t$ are the same.
Here, explicit value of the $S^{{\rm (cp)}}_1$  needs not to be known.
Instead, $v_{t p,\, dd}^{({\rm cp})} \,S_1^{\rm (cp)}$ and 
$v_{^3{\rm He} n,\, dd}^{({\rm cp})} \,S_1^{\rm (cp)}$ 
in Eqs.~(3.9) and (3.10)  are considered as adjustable parameters in the present 
$T$-matrix calculation of reactions (1.1) and (1.2);
then, the same parameters are used in the calculation of the $T$-matrix elements 
in the studies of reactions (1.3) and (1.4). 

The $p$-wave cross sections of the rearrangement reactions (1.1) and (1.2) are 
expressed in a simple closed form (3.9)--(3.12), that can reproduce observed data 
by tuning the potential parameters --- this is one of the  key findings of this study.

We determined  the potential parameters
$v_{^3{\rm He} n,\, dd}^{({\rm cp})}\, S_1^{(cp)},$ 
$v_{t p,\, dd}^{({\rm cp})}\, S_1^{(cp)}$, $\mu_1,\, \mu_4,$ and $\mu_7$, 
then use them to reproduce the five cases of the $p$-wave $S(E)$-factors 
$S_{\!dd\to^3{\rm He}n}(E)$ and $S_{\!dd\to tp}(E)$ in Fig.~\ref{fig:8line-sfactor}. 
Here,  $\mu_4= \mu_7$ is assumed.
We selected four sets of the coupling potential parameters, as listed 
in \mbox{Table~\ref{tab:vdtan},} for each of the five optical-potentials A to 
E (Table I), with the imaginary parts omitted.
Sets A1-A4 are obtained using the potential A, and similarly for B to E.
The strengths $v_{^3{\rm He} n, dd}^{({\rm cp})}$ and 
$v_{tp, dd}^{({\rm cp})}$ are for the case of Tumino+ (Arai+), 
while those for Angulo+, Nebia+, and Solovyev are left unwritten for simplicity.

\begin{table}
\caption{
Parameters of the \mbox{$dd$-$^3$He$n$} and 
$dd$-$tp$ coupling potentials in Eq. (3.5) and (3.6). $\mu_7=\mu_4$ is assumed.
Sets A1-A4 are determined using the optical-potential A \mbox{(Table I)} 
with the imaginary part omitted; similarly for B to E. The 
strengths $v_{^3{\rm He} n, dd}^{({\rm cp})}\,S_1^{({\rm cp})}$ and 
$v_{tp, dd}^{({\rm cp})}\,S_1^{({\rm cp})}$  are for the case of 
\mbox{Tumino+} (Nebia+), while those for Angulo+, Arai+ and Solovyev are 
not written to prevent complexity.}
\label{tab:vdtan}
  \begin{tabular}{ccccc}
\noalign{\vskip 0.1 true cm}
    \hline
    \hline
\noalign{\vskip 0.1 true cm}
Pot.   & \; $v_{^3{\rm He} n, dd}^{({\rm cp})}\,S_1^{({\rm cp})}$  
& $v_{tp, dd}^{({\rm cp})}\,S_1^{({\rm cp})}$& 
  $\mu^{-1/2}_1$  &   $\mu^{-1/2}_{4\,(7)}$ \\
\noalign{\vskip 0.1 true cm}
 Set&  (MeV fm$^{-5}$)  &  (MeV fm$^{-5}$)&  (fm)  & 
  (fm) \\
\noalign{\vskip 0.1 true cm}
\noalign{\vskip 0.0 true cm}
    \hline
\noalign{\vskip 0.1 true cm}
 A1   & \;\; 0.2579 \;\;(0.2942) \;\; &\;\; 0.2307\;\; (0.2136)\;\; & 3.5   & 2.0  \\
\noalign{\vskip 0.05 true cm}
 A2   & \;\;0.0600\;\; (0.0655)\;\;  &\;\; 0.0582\;\; (0.0515)\;\; & 2.0   & 4.0       \\
\noalign{\vskip 0.05 true cm}
 A3   & \;\;0.0022\;\; (0.0019)\;\;  &\;\; 0.0022\;\; (0.0016)\;\; & 5.5   & 5.5       \\
\noalign{\vskip 0.05 true cm}
 A4   & \;\;0.0086\;\; (0.0085)\;\;  &\;\; 0.0078\;\; (0.0063)\;\; & 6.0   & 3.0       \\
\noalign{\vskip 0.05true cm}
\hline
\noalign{\vskip 0.05 true cm}
\noalign{\vskip 0.05 true cm}
 B1   & \;\;0.0400\;\; (0.0439)\;\;  &\;\; 0.0410\;\; (0.0367)\;\; & 2.0   & 5.0       \\
\noalign{\vskip 0.05 true cm}
 B2   & \;\;0.0291\;\; (0.0310)\;\;  &\;\; 0.0275\;\; (0.0244)\;\; & 3.0   & 4.0       \\
\noalign{\vskip 0.05 true cm}
 B3   & \;\;0.0134\;\; (0.0165)\;\;  &\;\; 0.0141\;\; (0.0142)\;\; & 4.5   & 5.5       \\
\noalign{\vskip 0.05 true cm}
 B4   & \;\;0.0122\;\; (0.0138)\;\;  &\;\; 0.0112\;\; (0.0103)\;\; & 5.5   & 3.0       \\
\noalign{\vskip 0.05 true cm}
\hline
\noalign{\vskip 0.05true cm}
\noalign{\vskip 0.05 true cm}
 C1   & \;\;0.0805\;\; (0.0888)\;\;  &\;\; 0.0740\;\; (0.0660)\;\; & 3.0   & 3.0       \\
\noalign{\vskip 0.05 true cm}
 C2   & \;\;0.0434\;\; (0.0470)\;\;  &\;\; 0.0421\;\; (0.0378)\;\; & 2.0   & 4.5       \\
\noalign{\vskip 0.05 true cm}
 C3   & \;\;0.0019\;\; (0.0020)\;\;  &\;\; 0.0021\;\; (0.0018)\;\; & 5.0   & 6.0       \\
\noalign{\vskip 0.05 true cm}
 C4   & \;\;0.0030\;\; (0.0032)\;\;  &\;\; 0.0027\;\; (0.0025)\;\; & 6.0   & 3.5       \\
\noalign{\vskip 0.05 true cm}
\hline
\noalign{\vskip 0.05true cm}
\noalign{\vskip 0.05true cm}
 D1   & \;\;0.0028\;\; (0.0030)\;\;  &\;\; 0.0027\;\; (0.0024)\;\; & 5.0   & 4.5       \\
\noalign{\vskip 0.05 true cm}
 D2   & \;\;0.0140\;\; (0.0151)\;\;  &\;\; 0.0137\;\; (0.0123)\;\; & 4.0   & 4.5       \\
\noalign{\vskip 0.05 true cm}
 D3   & \;\;0.1737\;\; (0.1933)\;\;  &\;\; 0.1601\;\; (0.1482)\;\; & 2.0   & 3.5       \\
\noalign{\vskip 0.05 true cm}
 D4   & \;\;0.0108\;\; (0.0117)\;\;  &\;\; 0.0094\;\; (0.0086)\;\; & 6.0   & 2.5       \\
\noalign{\vskip 0.05 true cm}
\hline
\noalign{\vskip 0.05true cm}
\noalign{\vskip 0.05 true cm}
 E1   & \;\;0.0741\;\; (0.0767)\;\;  &\;\; 0.0658\;\; (0.0557)\;\; & 5.5   & 2.0       \\
\noalign{\vskip 0.05 true cm}
 E2   & \;\;0.4381\;\; (0.4951)\;\;  &\;\; 0.4010\;\; (0.3715)\;\; & 3.0   & 3.0       \\
\noalign{\vskip 0.05 true cm}
 E3   & \;\;0.0230\;\; (0.0219)\;\;  &\;\; 0.0230\;\; (0.0183)\;\; & 4.0   & 5.0       \\
\noalign{\vskip 0.05 true cm}
 E4   & \;\;0.1146\;\; (0.1255)\;\;  &\;\; 0.1109\;\; (0.0984)\;\; & 2.0   & 4.0       \\
\noalign{\vskip 0.05 true cm}
\hline
\hline
\noalign{\vskip 0.1 true cm}
\end{tabular}
\end{table}
%

\begin{figure}
\setlength{\abovecaptionskip}{0.cm}
\setlength{\belowcaptionskip}{-0.cm}
\centering
\includegraphics[width=0.46\textwidth]{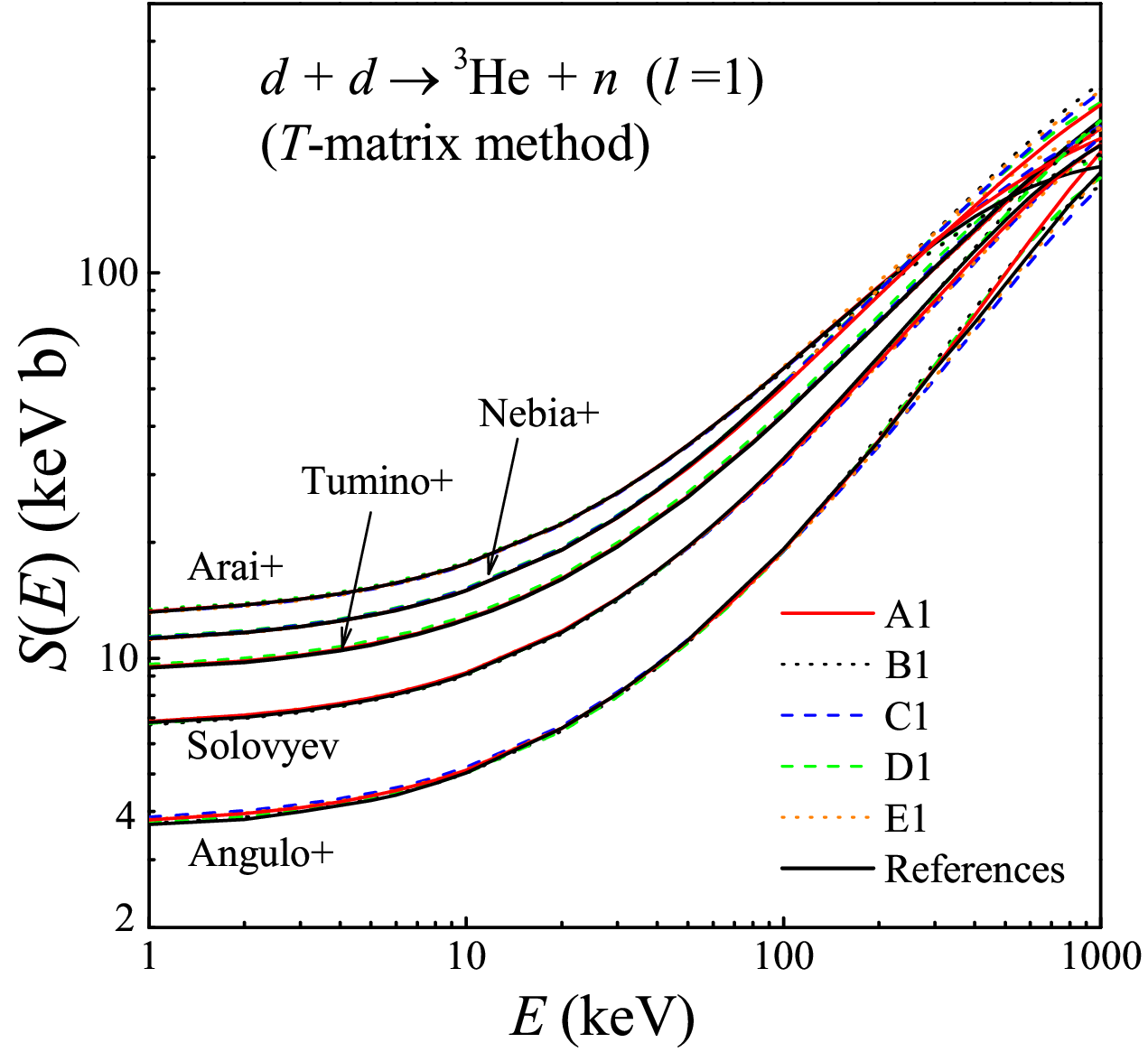}
\caption{
$p$-wave $S(E)$ factor of reaction (1.1), $S_{dd\to^3{\rm He}n}(E)$.
Five black lines are those reported by Angulo and Decouvemont~\cite{Angulo1998}, 
Nebia {\it et al.}~\cite{Nebia2002},
Arai {\it et al.}~\cite{Arai2011}, Tumino {\it et al.}~\cite{Tumino2014}, 
and Solovyev~\cite{Solovyev2024}. 
Lines A1-E1 closely reproducing  each black line are derived by the present 
$T$-matrix calculation using the $dd-^3$He$n$ coupling potentials A1-E1 
listed in \mbox{Table~\ref{tab:vdtan}};
use of the other coupling potential Ai-Ei ($i=2-4$) give similar results.
}
\label{fig:newT-S-hn}
\end{figure}

\begin{figure}
\setlength{\abovecaptionskip}{0.cm}
\setlength{\belowcaptionskip}{-0.cm}
\centering
\includegraphics[width=0.46\textwidth]{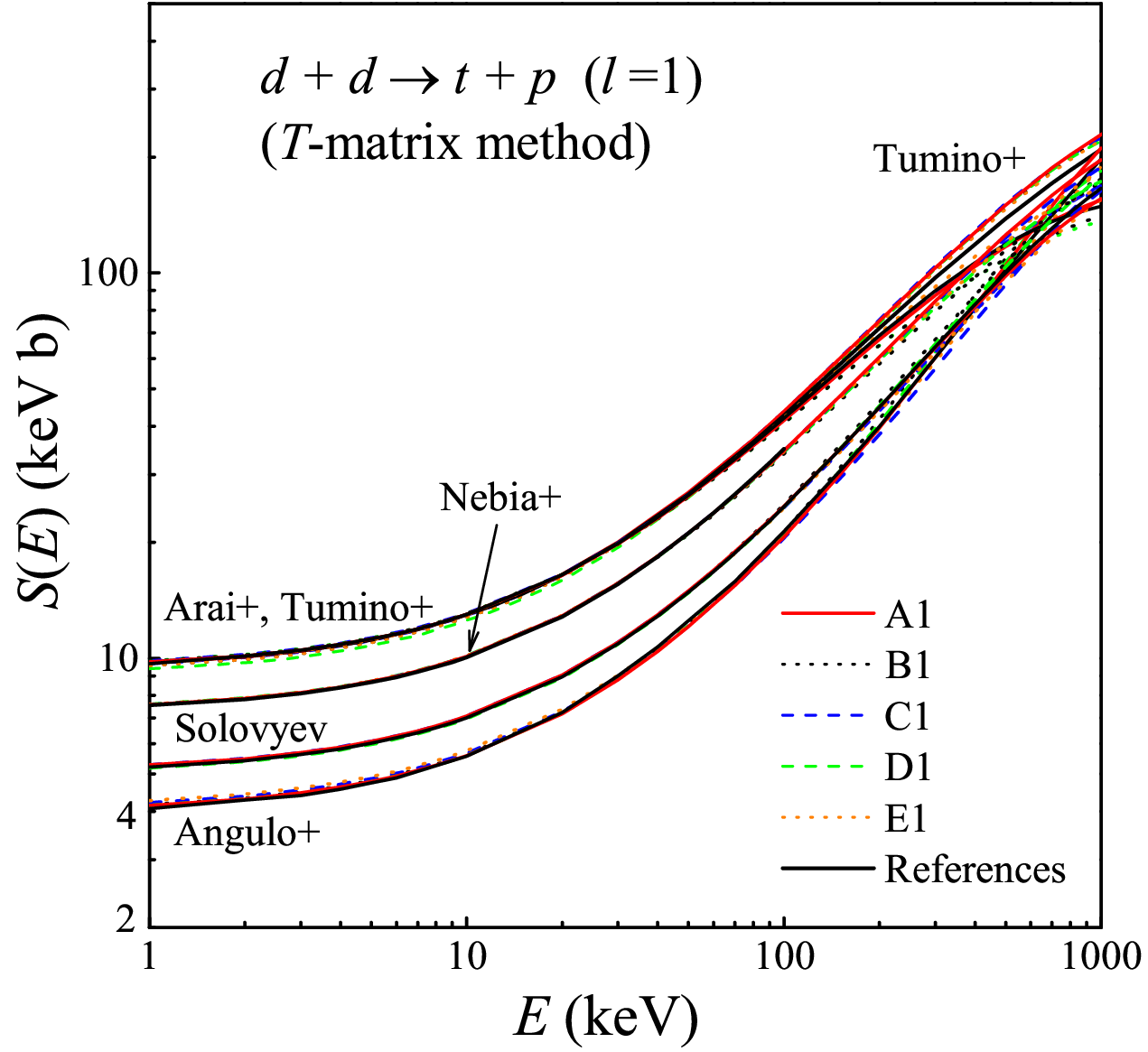}
\caption{
$p$-wave $S(E)$ factor of reaction (1.2), $S_{dd\to tp}(E)$.
Same meaning for lines as in Fig.~\ref{fig:newT-S-hn}.
}
\label{fig:newT-S-tp}
\end{figure}

In Figs. \ref{fig:newT-S-hn} and \ref{fig:newT-S-tp}, respectively, 
the calculated  $S_{\!dd\to^3{\rm He}n}(E)$ and $S_{\!dd\to tp}(E)$
using the five potential sets A1 to E1 are compared, in good agreement, 
with the five cases of $S(E)$ factors by 
Refs.~\cite{Angulo1998,Nebia2002,Arai2011,Tumino2014,Solovyev2024}
in black lines.
Use of the other 16 sets of the parameters in Table~\ref{tab:vdtan}
yield similar agreement. These coupling potentials are used in the following sections.



\section{Fusion rate of \boldmath{\lowercase{$dd\mu$}} molecule (\lowercase{ii}): \\
\vskip 0.05cm
     \boldmath{$T$}-matrix model on channels $\lowercase{c=}$ 5 and 8}

In this section, we calculate the fusion rates of the reactions,
\setcounter{equation}{2} 
\begin{equation*}
 (dd\mu^-)_{J=1,v}  
\stackrel{\lambda_{J=1,v}^{(^3{\rm He}n\mu)}}{\longrightarrow} 
  \!\! \begin{cases}
     \:  ^3{\rm He} + n + \mu^- +4.03\, \mbox{MeV},     \hskip 0.24cm (4.1a)\\
     \:  (^3{\rm He}\mu^-)_{nl} + n +4.03\, \mbox{MeV}, \hskip 0.17cm (4.1b)
   \end{cases}
\end{equation*}
\vskip -0.3cm
\begin{equation*}
 (dd\mu^-)_{J=1,v} \:\; \stackrel{\lambda_{J=1,v}^{(tp\mu)}}{\longrightarrow} \:\;
 \!\!  \begin{cases}
     \:  t + p + \mu^- +3.27\, \mbox{MeV},      \hskip 0.75cm (4.2a)\\
     \:  (t\mu^-)_{nl} + p +3.27\, \mbox{MeV},  \hskip 0.69cm (4.2b)\\
     \:  (p\mu^-)_{nl} + t +3.27\, \mbox{MeV},  \hskip 0.69cm (4.2c)
   \end{cases}
\end{equation*}
employing {\it method} ii);
namely, using the tractable three-body $T$-matrix model~\cite{Wu2024} 
taking channel $c=5$ and 8 (Fig.~\ref{fig:3body-jacobi}) for the description 
of the outgoing particles.
To formulate the fusion rate and the $T$-matrix of those reactions, 
we modify Eqs.~(4.5), (4.6), (4.9), and (4.10) in Ref.~\cite{Wu2024}.
Interactions that are determined in the previous sections (cf. Tables I and V) 
are used to reproduce the \mbox{$p$-wave} 
$S(E)$ factors in Fig.~\ref{fig:8line-sfactor}.

In order to treat the transition to the  three-body continuum channels (4.1a) 
and (4.2a), we employ the continuum-discretization method (cf. Sec.~IV of 
Ref.~\cite{Wu2024}) that was utilized by one of the present authors (M.K.) and 
collaborators for developing the CDCC (Continuum-Discretized Coupled-Channel) 
method for few-body reactions~\cite{Kamimura86,Austern,Yahiro12}.

\begin{figure}
\begin{center}
\epsfig{file=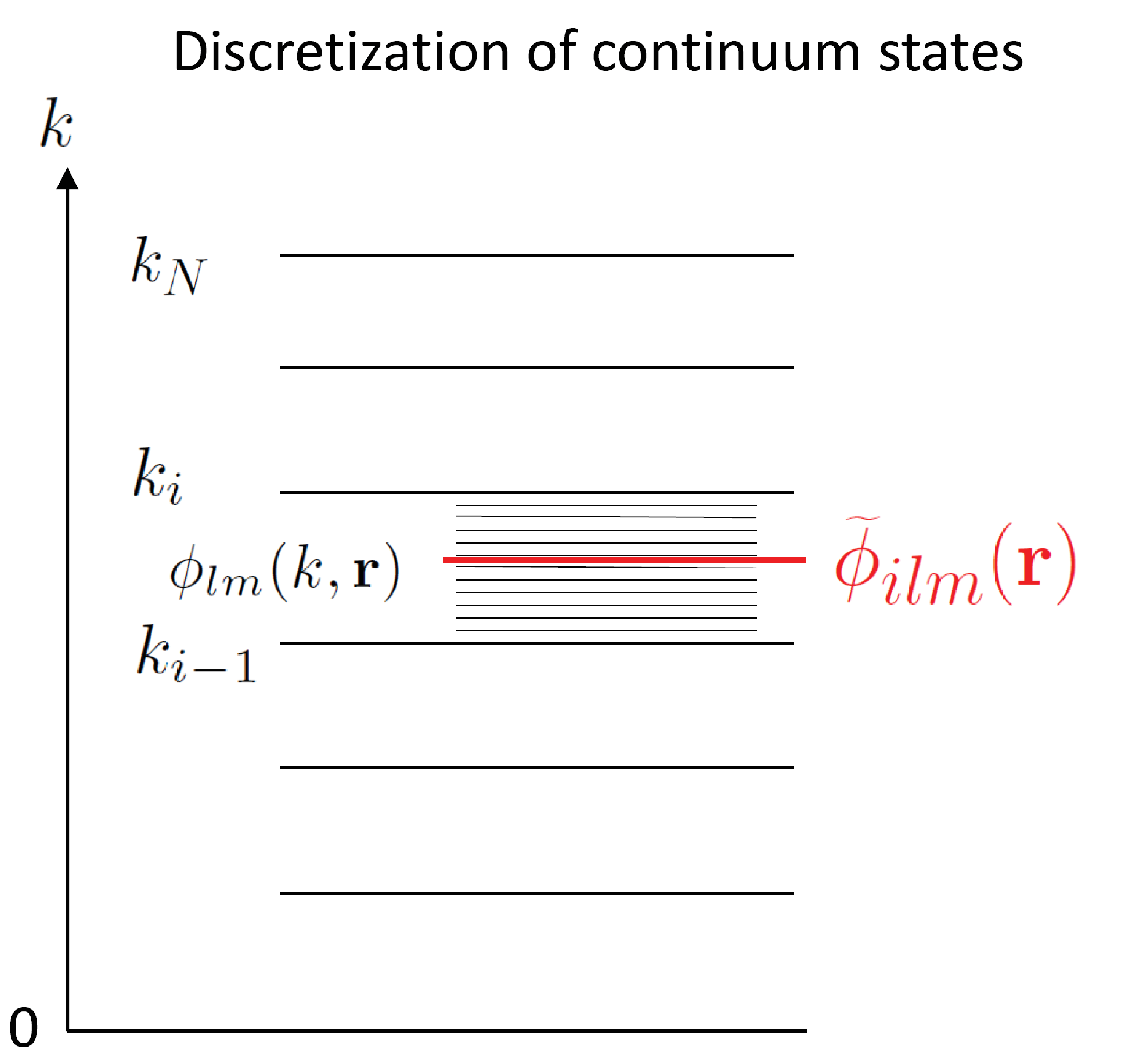,width=6.0cm,height=4.8cm}
\end{center}
\vskip -0.5cm
\caption{Schematic illustration of Eq.~(4.4) to construct the 
continuum-discretized wave function ${\widetilde \phi}_{i l m}({\bf r})$ 
by averaging the continuum wave functions $\phi_{lm}(k,{\bf r})$
in each momentum bin $\Delta k_i=k_i-k_{i-1}$. 
}
\label{fig:discretization}
\end{figure}

We discretize the reaction (4.1a) as $(i=1 - N)$,
\begin{eqnarray}
  (dd\mu)_{J=1,v} \to  (^3{\rm He}\mu)_{il} 
 + n  +4.03  \,\mbox{MeV},  \;\; 
\end{eqnarray}
where, as seen in Fig.~\ref{fig:discretization}, the $k$-momentum continuum states  
$\{ \phi_{lm}(k,{\bf r}_5), k=0 - k_N \}$ of the $^3{\rm He}$-$\mu$ subsystem
are discretized into the orthonormalized states 
\mbox{$\{ {\widetilde \phi}_{i l m}({\bf r}_5), i=1 - N \}$} by
\begin{eqnarray}
 \!\!\!\!  \!\!\!\! &&{\widetilde \phi}_{i l m}({\bf r}_5)
 =\frac{1}{\sqrt{\mathit{\Delta}k_i}}
\int_{k_{i-1}}^{k_{i}} \!\! \phi_{lm}(k,{\bf r}_5)\, dk, \;  
  \\
\label{eq:bin}
 \!\!\!\!  \!\!\!\! &&{\widetilde \varepsilon}_{i}
=\frac{\hbar^2}{2\mu_{r_5}} {\widetilde k}_i^{\,2},  \qquad
    {\widetilde k_i}^{\,2}=\Big(\frac{k_i+k_{i-1}}{2}\Big)^2 
+ \frac{\mathit{\Delta}k_i^2}{12} \;  
\end{eqnarray}
with  ${\widetilde \varepsilon}_{i}$ and ${\widetilde k_i}$ being the average energy and 
momentum of ${\widetilde \phi}_{i l m}({\bf r}_5)$. 
Similarly to the $^4{\rm He}$-$\mu$ case in the $(dt\mu)$ molecule~\cite{Wu2024},
we consider $N=200$ for $l=0$ to 15, and the maximum momentum
$\hbar k_N=10.0 $\, MeV/$c \,
({\widetilde \varepsilon}_{N}=487$ keV) with the constant $\Delta k_i$.
This is precise enough to derive a continuous function of $k$ for the momentum spectrum 
of the $^3{\rm He}$-$\mu$ continuum (cf. Eq.~(\ref{eq:k-convergence}) below).

%


The $T$-matrix elements and transition rates to the continuum-discretized channel 
$(^3{\rm He}\mu)_{il} + n$ on $c=5$ are described  as follows by modifying  
Eqs.~(4.5) and (4.10) of Ref.~\cite{Wu2024},
\begin{eqnarray}
&&\!\!\!\!\!\!\!\!
{\widetilde T}^{(c=5)}_{Jv,ilm}({\widetilde {\bf K}}_i) = 
    \langle e^{ i {\widetilde {\bf K}}_i \cdot {\bf R}_5 } \,
      {\widetilde \phi}_{ilm}({\bf r}_5)
    |\,  {\cal V}^{\rm (cp)}_{\!^3{\rm He} n, dd} \,| \,
 \Phi^{\rm (opt)}_{JM, v}(dd\mu)  \,\rangle,   \nonumber \\ \\
 && \!\!\!\!\!\!\!\!
{\widetilde r}_{Jv,il}^{\,(c=5)} ={v}^{(5)}_{il}
  \!  \left( \frac{\mu_{R_5}}{2 \pi \hbar^2} \right)^{\!2} 
     \! \big| S^{\rm (cp)}_1 \, \big|^2
   \sum _{m}\! \int \big| {T}^{(c=5)}_{Jv,ilm}
   ({\widetilde {\bf K}}_i) \,\big|^2   {\rm d}{\widehat {\widetilde {\bf K}}_i},
    \nonumber \\  
\label{eq:T5-cont1}
\end{eqnarray}
where the magnitude of the wave number  
${\widetilde {\bf K}}_i$ of the plane wave along ${\bf R}_5$ 
is derived from energy conservation as,
\begin{eqnarray}
\hbar^2{\widetilde K}_i^2/2\mu_{R_5}  + {\widetilde \varepsilon}_{i}
 = E_{Jv}^{\rm (real)} + 3.27 \, \mbox{MeV},  
\label{eq:E-conservation-2}
\end{eqnarray}
and similarly for $K_n$ below.

The transition to the $(^3{\rm He}\mu)_{nl} + n$ channel in the reaction (4.1b)
is represented by modifying  Eqs. (4.6) and (4.9) of Ref.~\cite{Wu2024} as,
\begin{eqnarray}
&&\!\!\!\!\!\!\!\!
T^{(c=5)}_{Jv,nlm}({\bf K}_n) = 
    \langle e^{ i {\bf K}_n \cdot {\bf R}_5 } 
       \phi_{nlm}({\bf r}_5)
    |  {\cal V}^{\rm (cp)}_{\!^3{\rm He} n, dd} | 
 \Phi^{\rm (opt)}_{JM, v}(dd\mu)  \rangle, \!\!\!\! \nonumber \\ \\
 && \!\!\!\!\!\!\!\!
 r_{Jv,nl}^{\,(c=5)} ={v}^{(5)}_{nl}
  \!  \left( \frac{\mu_{R_5}}{2 \pi \hbar^2} \right)^{\!2} 
     \! \big| S^{\rm (cp)}_1 \, \big|^2
   \sum _{m}\! \int \! \big| {T}^{(c=5)}_{Jv,nlm}
   ({\bf K}_n) \big|^2   {\rm d}{\widehat {\bf K}_n}.
    \nonumber \\  
\label{eq:T5-bound}
\end{eqnarray}
In the above definition of $T$-matrix elements in Eqs.~(4.6) and (4.9),
the spin part is not included but represented by the factor 
$\big| S^{\rm (cp)}_1 \, \big|^2$ in Eqs.~(4.7) and (4.10),
since the coupling interaction 
$V^{\rm (cp)}_{^3{\rm He} n, dd}$ does not depend on spins (cf. Eq.~(3.14)).

The {\it ket} vector amplitude of Eqs.~(4.6) and (4.9), 
namely $\Phi^{\rm (opt)}_{JM,v}(dd\mu)$ obtained by Eq.~(2.10) 
using the $d$-$d$ optical-potential, is employed in place of the {\it ket}-vector 
amplitudes of the CC-solution (5.2)-(5.3) in Ref.~\cite{Kamimura2023}. 
In $\Phi^{\rm (opt)}_{JM,v}(dd\mu)$, the effects of the
outgoing $^3{\rm He} n \mu$ and $t p \mu$ channels are reflected to a considerable 
extent through the imaginary part $W^{\rm (N)}_{dd}$ of the optical-potential.

The reactions to the $tp\mu$ system in Eqs.~(4.2a)
and (4.2b) are formulated similarly as above by changing
channel $c=5$ to $c=8$. 
We first discretize the $t + p + \mu$ channel (4.2a) as ($i=1-N$)
\begin{eqnarray}
  (dd\mu)_{J=1,v} \to  (t\mu)_{il}  + p  +3.27 \,\mbox{MeV}.  
\end{eqnarray}
The $T$ matrix and the reaction rate to the above continuum-discretized channel are 
described  in the same way by,
\begin{eqnarray}
&&\!\!\!\!\!\!\!\!
{\widetilde T}^{(c=8)}_{Jv,ilm}({\widetilde {\bf K}}_i) = 
    \langle e^{ i {\widetilde {\bf K}}_i \cdot {\bf R}_8 } \,
      {\widetilde \phi}_{ilm}({\bf r}_8)
    |\,  {\cal V}^{\rm (cp)}_{tp, dd} \,| \,
 \Phi^{\rm (opt)}_{JM, v}(dd\mu)  \,\rangle,   \nonumber \\ \\
 && \!\!\!\!\!\!\!\!
{\widetilde r}_{Jv,il}^{\,(c=8)} ={v}^{(8)}_{il}
  \!  \left( \frac{\mu_{R_8}}{2 \pi \hbar^2} \right)^{\!2} 
     \! \big| S^{\rm (cp)}_1 \, \big|^2
   \sum _{m}\! \int \big| {T}^{(c=8)}_{Jv,ilm}
   ({\widetilde {\bf K}}_i) \,\big|^2   {\rm d}{\widehat {\widetilde {\bf K}}_i}.
    \nonumber \\  
\label{eq:T8-cont1}
\end{eqnarray}
The transition to the  $(t\mu)_{nl} + p$ channel in (4.2b) is given as 
\begin{eqnarray}
&&\!\!\!\!\!\!\!\!
T^{(c=8)}_{Jv,nlm}({\bf K}_n) = 
    \langle e^{ i {\bf K}_n \cdot {\bf R}_8 } 
       \phi_{nlm}({\bf r}_8)
    |  {\cal V}^{\rm (cp)}_{tp, dd} | 
 \Phi^{\rm (opt)}_{JM, v}(dd\mu)  \rangle, \!\!\!\! \nonumber \\ \\
 && \!\!\!\!\!\!\!\!
 r_{Jv,nl}^{\,(c=8)} ={v}^{(8)}_{nl}
  \!  \left( \frac{\mu_{R_8}}{2 \pi \hbar^2} \right)^{\!2} 
     \! \big| S^{\rm (cp)}_1 \, \big|^2
   \sum _{m}\! \int \! \big| {T}^{(c=8)}_{Jv,nlm}
   ({\bf K}_n) \big|^2   {\rm d}{\widehat {\bf K}_n}.
    \nonumber \\  
\label{eq:T8-bound}
\end{eqnarray}
The reaction (4.2c), where the outgoing particles are on the $c=9$ channel, can be 
described using Eqs.~(4.12) to (4.15) with changing `8' to `9'.

By the way, the $T$-matrix elements (4.6), (4.9), (4.12) and (4.14)
require multiple integrals.
The following treatment will be useful in the actual 
calculations with representing $\Phi^{\rm (opt)}_{JM, v}(dd\mu)$ as 
$\Phi^{\rm (opt)}_{JM, v}({\bf r}_1, {\bf R}_1) $,
\begin{eqnarray}
{\cal V}^{\rm (cp)}_{\!^3{\rm He}n, dd}\!\!\!\!\!\!&&\!\!\!\! (dd\mu)\,
\Phi^{\rm (opt)}_{JM, v}(dd\mu) \nonumber \\
&=\!\!& \int \! V^{\rm (cp)}_{\!^3{\rm He} n, dd}({\bf r}_4, {\bf r}_1) 
  \, \Phi^{\rm (opt)}_{JM, v}({\bf r}_1, {\bf R}_1) \,{\rm d} {\bf r}_1 \, 
   \qquad \nonumber  \\
&=\!& v^{\rm (cp)}_{^3{\rm He}n, dd} \, A(r_4) \, B_{Jv}(R_4)  
 \, \Big[ Y_1({\hat {\bf r}}_4)\,Y_1({\hat {\bf R}}_4)\Big]_0, \qquad \quad \\
\!\!\!\!{\cal V}^{\rm (cp)}_{\!tp, dd}\!\!\!\!\!\!&&\!\!\!\!(dd\mu) \,
 \Phi^{\rm (opt)}_{JM, v}(dd\mu)  \nonumber \\
&=\!\!& \int \! V^{\rm (cp)}_{\!tp, dd}({\bf r}_7, {\bf r}_1) 
  \, \Phi^{\rm (opt)}_{JM, v}({\bf r}_1, {\bf R}_1) \,{\rm d} {\bf r}_1 \, 
   \qquad \nonumber  \\
&=\!& v^{\rm (cp)}_{tp, dd} \, A(r_7) \, B_{Jv}(R_7)  
 \, \Big[ Y_1({\hat {\bf r}}_7)\,Y_1({\hat {\bf R}}_7)\Big]_0,
\end{eqnarray}
where we take ${\bf R}_1={\bf R}_4={\bf R}_7 $ (cf. Fig.~\ref{fig:3body-jacobi}).
We can then transform the Jacobi coordinates $({\bf r}_4, {\bf R}_4)$ to  
$({\bf r}_5, {\bf R}_5)$ in Eq.~(4.16), and $({\bf r}_7, {\bf R}_7)$ to  
$({\bf r}_8, {\bf R}_8)$ in Eq.~(4.17).

Summation over $n$ for the reaction rates $r^{(c=5)}_{Jv,nl}$ of Eq.~(4.10) 
yields the reaction rates $r^{(c=5)}_{Jv,l}{\rm (bound)}$ for the bound states,
%
\begin{eqnarray}
r^{(^3{\rm He}n\mu)}_{Jv,\,l}{\rm (bound)}=\!\! 
\sum_n r^{(c=5)}_{Jv,\,nl}\,.
\label{eq:rate-bound-3Hen}
\end{eqnarray}
As for the continuum states, we transform the summation  
$\sum\nolimits_{i} {\widetilde r}_{Jv,il}^{(c=5)}$ 
into the integration of a smooth continuum function 
$r_{Jv,l}^{(c=5)}(k)$ of $k$ as,\footnote{
A test of this $\mathit{\Delta}k \to 0$ process is well explained in the review papers of 
the CDCC method~\cite{Kamimura86,Austern,Yahiro12}.}
\begin{eqnarray}
\sum_{i=1}^{K_N} {\widetilde r}_{Jv,il}^{(c=5)}
 \! =\!\sum_{i=1}^{K_N} \Big( \frac{{\widetilde r}_{Jv,il}^{(c=5)}}
{\mathit{\Delta}k} \Big)\!
 \mathit{\Delta}k
 \stackrel{\mathit{\Delta}k \to 0}{\longrightarrow} 
 \!\!\! \int_0^{\,k_N}\!\!\! r_{Jv,l}^{(c=5)}(k)\, {\rm d} k.  \quad \;\;
\label{eq:k-convergence}
\end{eqnarray}  
\noindent
Then, the sum over the quantum number $i$ for ${\widetilde r}^{\,(c)}_{Jv,il}$ yields 
the total reaction rates $r^{(c)}_{Jv,l}{\rm (cont.)}$ for the continuum states,
\begin{eqnarray}
r^{(^3{\rm He}n\mu)}_{Jv,\,l}{\rm (cont.)}=  
\! \int_0^{\,k_N}\! r_{Jv,l}^{(c=5)}(k)\, {\rm d} k 
\label{eq:rate-bound-3Hen}
\end{eqnarray}
and similarly for the $(t\mu)$-$n$ system on the $c=8$ channel.
Summing up over $l$, we have the total reaction rates to the 
$^3{\rm He}$-$\mu$ bound  and continuum states,
\begin{eqnarray}
 \lambda_{Jv}^{(^3{\rm He}n\mu)}({\rm bound})
&=&\sum_{l=0}^{5}\:  r^{(^3{\rm He}n\mu)}_{Jv,\,l}{\rm (bound)}, \qquad  \\
 \lambda_{Jv}^{(^3{\rm He}n\mu)}({\rm cont.})
&=&\sum_{l=0}^{20}\:  r^{(^3{\rm He}n\mu)}_{Jv,\,l}{\rm (cont.)}, \qquad
\label{eq:lambda-cont}
\end{eqnarray}
and similarly for the $tp\mu$ system.


The sum of these transition rates,
\begin{eqnarray} 
\lambda_{Jv}^{({\rm ^3He}n\mu)}&\!=\!& \lambda_{Jv}^{(^3{\rm He}n\mu)}({\rm bound})
 + \lambda_{Jv}^{(^3{\rm He}n\mu)}({\rm cont.}), \quad \qquad \\
\lambda_{Jv}^{(tp\mu)}&\!=\!& \lambda_{Jv}^{(tp\mu)}({\rm bound})
 + \lambda_{Jv}^{(tp\mu)}({\rm cont.}),  \\
\lambda_{Jv}&\!=\!& \lambda_{Jv}^{(^3{\rm He}n\mu)}
 + \lambda_{Jv}^{(tp\mu)},  
\label{eq:fusion-rate-muon}
\end{eqnarray}
%
are the fusion rates of the $(dd\mu)_{Jv}$ molecule, using the 
\mbox{$T$-matrix} based on channels $c=5$ and 8, respectively.

\begin{figure}
\setlength{\abovecaptionskip}{0.cm}
\setlength{\belowcaptionskip}{-0.cm}
\centering
\includegraphics[width=0.43\textwidth]{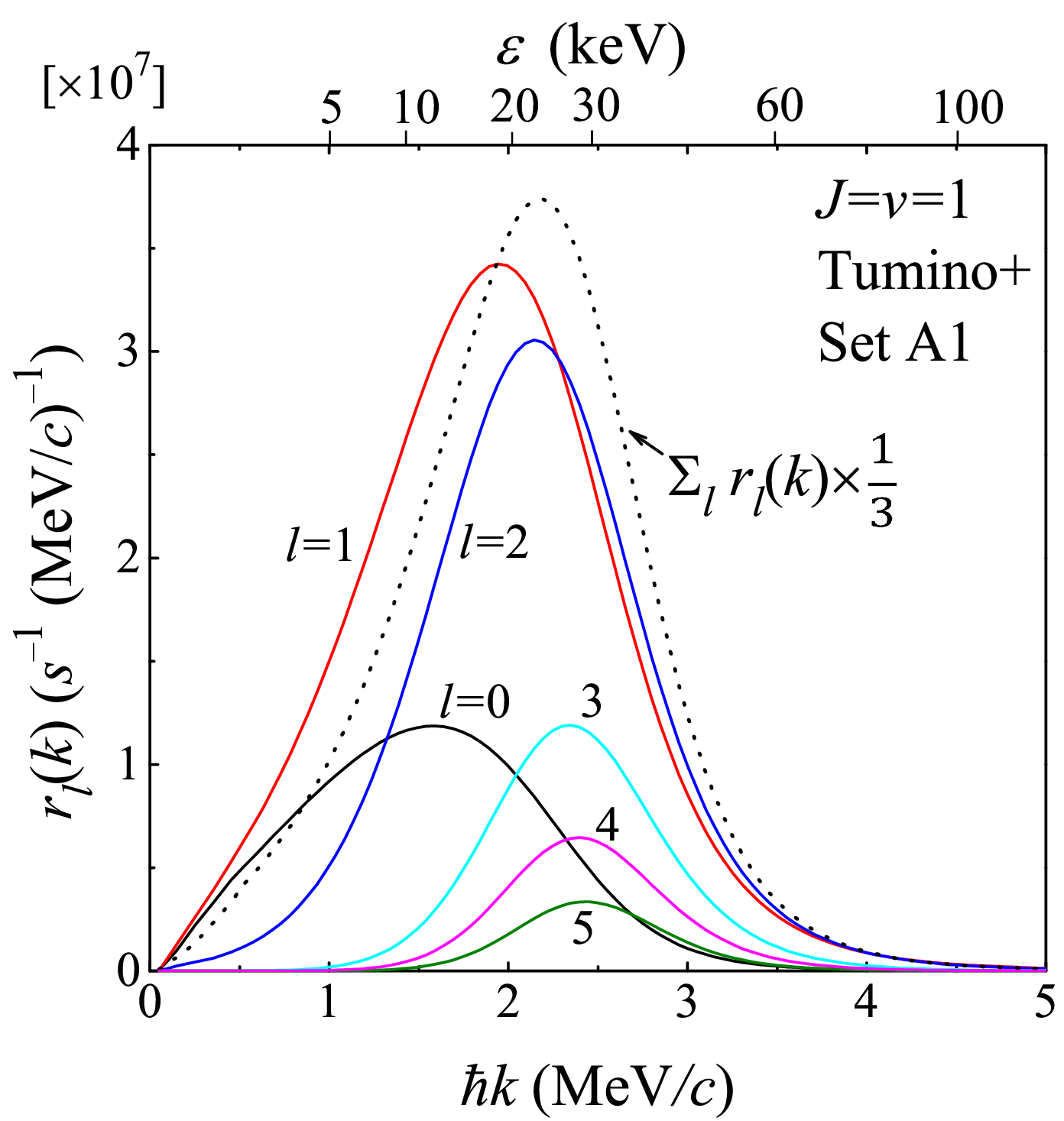}
\caption{
Calculated reaction rates ${r}_{Jv,l}^{(c=5)}(k)$ in Eq.~(4.20) 
of the ($dd\mu)_{J=v=1}$ molecule decaying to the $^3{\rm He}$-$\mu$ 
continuum states, with angular momentum $l$. 
Potential set A1 is used for  the $S(E)$ factor Tumino+ in Table~\ref{tab:vdtan}; 
uses of the other potential cases give similar results. 
The black dotted curve represents 
$\Sigma_{l=0}^{15} r_{Jv,l}^{(c=5)}(k)$  multiplied by $\tfrac{1}{3}$.
The reaction rates ${r}_{J=1,v=0,\,l}^{(c=5)}(k)$ decaying from ($dd\mu)_{J=1,v=0}$ 
exhibit almost the same behavior as the above curves multiplied \mbox{by 3.1.}
}
\label{fig:rlk-J1V1}
\end{figure}

The calculated continuum reaction rates ${r}_{J=v=1,\,l}^{(c=5)}(k)$ in 
Eq.~(4.20) are shown in Fig.~\ref{fig:rlk-J1V1},
for the angular momenta $l$ between $^3{\rm He}$ and $\mu$
using potential set A1 in Table~\ref{tab:vdtan}; uses of the other potential cases 
give similar results.
We see that the peak position of the dotted curve is at 
$\hbar k \approx 2.2 \, {\rm MeV}/c$ ($\varepsilon \approx 23$ keV).
This is understood as follows: 
with the kinetic energy 0.82 MeV (with speed 
$v_{^3{\rm He}}/c=0.024$) after the fusion,
the $^3{\rm He}$ particle  escapes
from the muon cloud, which has approximately the $(^4{\rm He} \mu)_{1s}$ wave 
function of ${\bf R}_4$.
Conversely, the muon cloud is moving with respect to the $^3{\rm He}$ particle 
with the same speed $v_{^3{\rm He}}/c$, namely $\hbar k \approx 2.5\, {\rm MeV}/c$, 
which is close to the peak \mbox{position.} 
The width of the peak of the dotted curve, corresponds to the width 
$(\approx 1.5 \,$MeV/$c$) 
of the momentum distribution of 
the muon $1s$ cloud. 

\begin{figure}
\setlength{\abovecaptionskip}{0.cm}
\setlength{\belowcaptionskip}{-0.cm}
\centering
\includegraphics[width=0.44\textwidth]{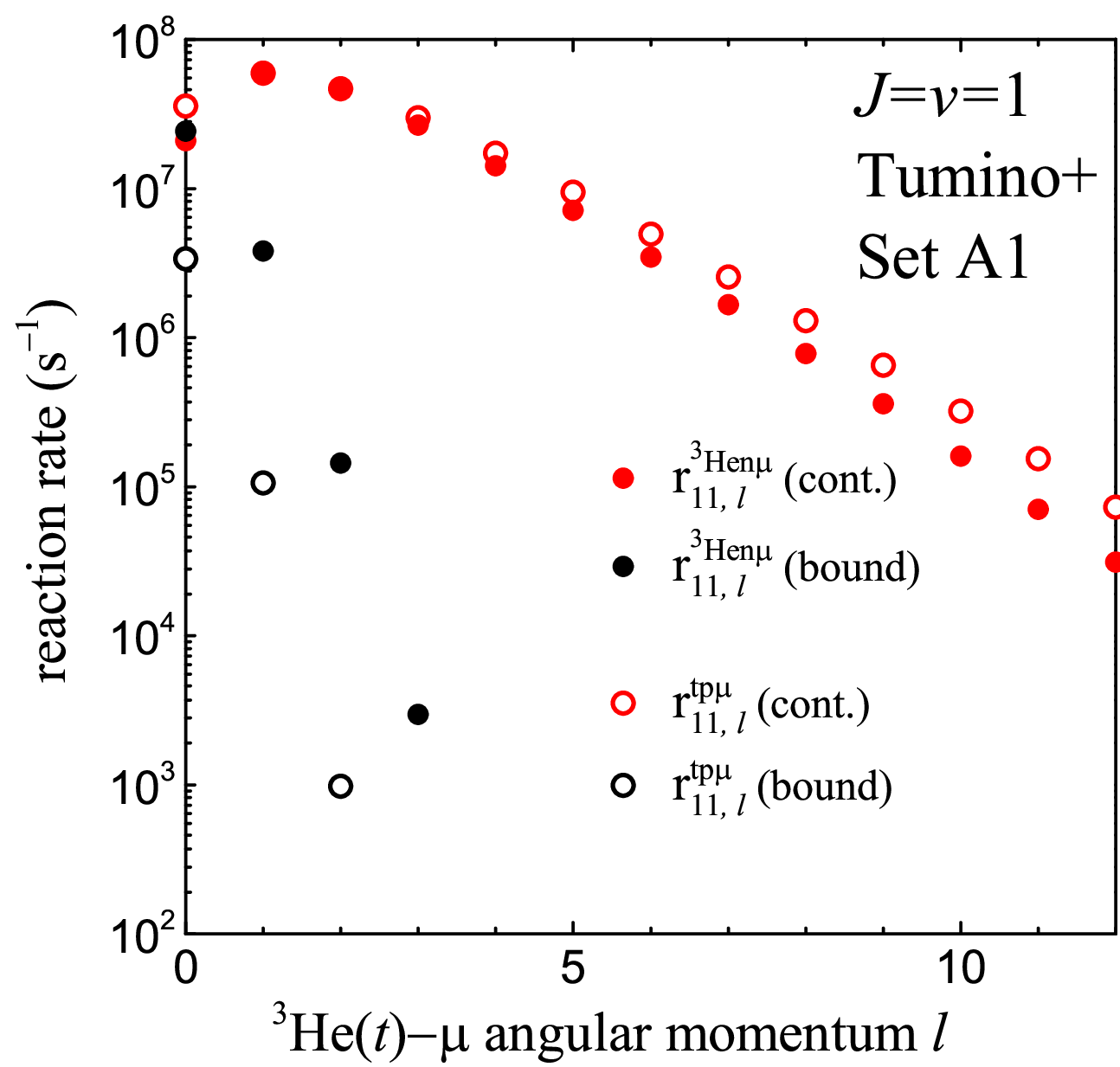}
\caption{
Calculated reaction rates  $r^{(^3{\rm He}n\mu)}_{Jv,l}{\rm (cont.)}$ 
in Eq.~(4.22) and $r^{(^3{\rm He}n\mu)}_{Jv,l}{\rm (bound.)}$ in Eq.~(4.21), 
decaying from the $(dd\mu)_{J=v=1}$ state to the $(^3{\rm He}n\mu)_l$ continuum 
and bound states derived on channel $c=5$; similarly for the $(t\mu)$-$p$ system 
on channel $c=8$. 
The potential set A1 is used for  the $S(E)$ factor Tumino+ in 
Table~\ref{tab:vdtan}; uses of the other potential cases give  similar results. 
The reaction rates decaying from ($dd\mu)_{J=1,v=0}$
exhibit almost the same behavior as the above circles multiplied by 3.1.}
\label{fig:rL-J1V1}
\end{figure}

Fig.~\ref{fig:rL-J1V1} illustrates how the reaction rates $r^{(^3{\rm He}n\mu)}_{J=v=1,\,l}{\rm (bound)}$ and
$r^{(^3{\rm He}n\mu)}_{J=v=1,\,l}{\rm (cont.)}$ 
in the R.H.S. of (4.21) and (4.22), respectively,  depend on the angular momentum $l=0$ to 12. 
The former rates decrease quickly with \mbox{increasing $l$}, whereas the latter change slowly. 
The ratio of these two rates is the essence of the initial $^3{\rm He}$-$\mu$ sticking probability, which will be discussed in the next section.
The reason why so many angular momenta $l$ appear in the reaction rates in Fig.~\ref{fig:rL-J1V1} is, \mbox{in the \mbox{$T$-matrix}} elements (4.6) and (4.9), the component ${\cal V}^{\rm (cp)}_{^3{\rm He}n, dd}\, \Phi^{\rm (opt)}_{JM, v}(dd\mu)$ is composed of very short-range functions of ${\bf r}_4$ and long-range functions of ${\bf R}_4$ (cf.~Eq.(4.16)). 
Therefore, many angular momenta $l$ are necessary to expand
this unique function of $({\bf r}_4, {\bf R}_4)$ 
in terms of the functions of the rearrangement Jacobi coordinates
$({\bf r}_5, {\bf R}_5)$.

\begin{table*}
\caption{
Calculated fusion rates
$\lambda_{Jv}^{(^3{\rm He}n\mu)}$, $\lambda_{Jv}^{(tp\mu)}$
and their sum $\lambda_{Jv}$ 
of the $(dd\mu)_{J,v}$ states, with $J=1, v=1$ and $0$, 
calculated on the channels $c=5$ and $8$ using the 20 potential sets A1 to 
E4 (cf. Table~\ref{tab:vdtan}).
All in unit of $10^8 {\rm s}^{-1}$.
}
\label{tab:vdtanV}
  \begin{tabular}{cccccccc}
\noalign{\vskip 0.1 true cm}
    \hline
    \hline
\noalign{\vskip 0.1 true cm}
$p$-wave & $c=5$  & $c=8$  & $c=5\& 8$   & & $c=5$ &  $c=8$
&  $c=5\& 8$   \\
\noalign{\vskip 0.02true cm}
\;\;\;\; $S(E)$ factor   \;\;\;\;  &  $ \;\; \lambda_{11}^{(^3{\rm He}n\mu)}$  \;\; 
& $\lambda_{11}^{(tp\mu)}$  \;\;  &  \;\; 
$\lambda_{11}$  \;\; & $   $  \;\; &  \;\; 
$\lambda_{10}^{(^3{\rm He}n\mu)}$  \;\; &  \;\; 
$\lambda_{10}^{(tp\mu)}$  \;\; &  \;\; $\lambda_{10}$  \;\; \\
\noalign{\vskip 0.1 true cm}
\noalign{\vskip 0.0 true cm}
    \hline
\noalign{\vskip 0.1 true cm}
\noalign{\vskip 0.05true cm}
Angulo+ 1998 &  0.85(3) \: &  0.94(3)\:\,  & \;\;\;1.8(1) \:\,& & 
             2.7(1) \: & 2.9(1) \: &  \:\:5.6(2) \:  \\
\noalign{\vskip 0.1 true cm}
Nebia+ 2002 &  2.5(1) \: &  1.7(1)\:\,  & \;\;\;4.2(1) \:\,& & 
             7.8(1) \: & 5.2(1) \: &  \:\:13.0(2) \:  \\
\noalign{\vskip 0.1 true cm}
Arai+ 2011 &  2.9(1) \: &  2.1(1)\:\,  & \;\; 5.1(1) \:\,& & 
             9.1(1) \: & 6.7(1) \: &  15.8(1) \:  \\
\noalign{\vskip 0.1 true cm}
Tumino+ 2014  &  2.1(1) \: &  2.1(1)\:\,  & \;\;4.2(1) \:\,& & 
             6.6(1) \: & 6.4(1) \: &  13.1(2) \:  \\
\noalign{\vskip 0.05true cm}
Solovyev 2024   &  1.5(1) \: &  1.2(1)\:\,  & \;\;2.7(1) \:\,& & 
             4.8(1) \: & 3.6(1) \: & \;8.4(2) \:  \\
\noalign{\vskip 0.05true cm}
\hline
\hline
\noalign{\vskip 0.1 true cm}
\end{tabular}
\end{table*}

Table \ref{tab:vdtanV} lists the fusion rates $\lambda_{Jv}^{({\rm ^3He}\mu)}$,
$\lambda_{Jv}^{(tp\mu)}$, and their sum $\lambda_{Jv}$
calculated on channel $c=5$ and 8 for the \mbox{$J=1$} states, with $v=1$ and 0 
using the 20 potential sets (cf. Table~\ref{tab:vdtan}).
The most important point in Table \ref{tab:vdtanV} is the fact that
the values of the $T$-matrix model results $\lambda_{11}$ and 
$\lambda_{10}$ agree respectively with the optical-potential model 
results $\lambda_{11}^{\rm (opt)}$ and
$\lambda_{10}^{\rm (opt)}$. 
This indicates the validity of the two models.
Comparison with the observed value of the fusion rate
will be made at the end of Sec. VI

%
\section{Muon sticking probability}

The initial $^3{\rm He}$-$\mu$ sticking probability, $\omega^{Jv}_d$, 
is defined as the probability of the muon being captured by a $^3{\rm He}$ particle after the $(dd\mu)_{J,v}$ fusion reaction (4.1)~\cite{Bogdanova1985}, which is expressed as, 
\begin{eqnarray}
\omega^{Jv}_{\rm d}=\frac{\lambda_{Jv}^{(^3{\rm He}n\mu)}(\rm bound)}  
                       {\lambda_{Jv}^{(^3{\rm He}n\mu)}(\rm bound)
                    +   \lambda_{{\it {Jv}}}^{(^3{\rm He}n\mu)}(\rm cont.)},
\label{eq:initial-stick}
\end{eqnarray}
employing the fusion rates (4.21) and (4.22) listed in \mbox{Table~VI} 
in the same manner as Eq.~(5.13) of Ref.~\cite{Kamimura2023} and Eq.~(4.14) 
of Ref.~\cite{Wu2024}.

It is to be stressed that, here we do not take the sudden approximations as 
usually employed in the literature, but use the {\it absolute} values of 
the above two fusion rates, as was done for the $(dt\mu)$ 
molecule~\cite{Kamimura2023,Wu2024}. 
We further emphasize that the fusion rates are calculated explicitly using 
the nuclear interactions that reproduce
the $p$-wave $S(E)$ factors of the reactions (1.1) and (1.2) 
in the broad energy region of $E \simeq 1$ keV to 
\mbox{1 MeV~\cite{Tumino2014}}, as illustrated in Figs.~4 and 5.

Take what is shown in Fig.~\ref{fig:rL-J1V1} as an example, after summing up 
over $l$, we have $\lambda^{(^3{\rm He}n\mu)}_{J=v=1}{\rm (bound)}
=0.2816\times 10^8 {\rm s}^{-1}$ and
$\lambda^{(^3{\rm He}n\mu)}_{J=v=1}{\rm (cont.)}=1.8215\times 10^8 {\rm s}^{-1}$, 
giving $\omega_{\rm d}^{11}=0.1339$. 
For all the 20 sets of the nuclear interactions, the $\omega^{Jv}_{\rm d}$ 
are summarized in Table~\ref{tab:vdtanVI}, with the average, 
\begin{eqnarray}
  && \qquad  \omega_{\rm d}^{11}= 0.133 \pm 0.001,   \\   
  && \qquad  \omega_{\rm d}^{10}= 0.133 \pm 0.001, 
\end{eqnarray}
close to those employing the sudden approximation:
$\omega_{\rm d}^{11} (\omega_{\rm }^{10})=0.133\footnote{
According to Ref.~\cite{Markushin1988},
$\omega_{\rm d}^{11}$ in Ref.~\cite{Bogdanova1985} was originally 0.133, 
but multiplied by a normalization coefficient, giving 0.137.}\, (0.132)$ 
by Bogdanova {\it et al.}~\cite{Bogdanova1985}, 
0.1308 (0.1356) by Hu and Kauffmann~\cite{Hu1987}, 
and 0.13401 (0.13429) by Haywood {\it et al.}~\cite{Haywood1991}.

It is interesting to see that while the values of the fusion rates 
in \mbox{Table~VI} are somewhat scattered between the five cases of 
the $S(E)$ factors, the values of the sticking probabilities in 
Table~\ref{tab:vdtanVI} are concentrated at 0.133. This is because 
the sticking probability is a `ratio' of the fusion rates as seen in Eq.~(5.1).

The latest observation of the \mbox{`effective'} sticking probability 
by Balin {\it et al.}~\cite{Balin2011} gave
$\omega_{\rm d}^{\rm eff}{\rm (exp)}=0.1224\, (6)$
for gas density $\varphi=0.0837$.
$\omega_{\rm d}^{\rm eff}{\rm (exp)}$ corresponds to the theoretically obtained
initial sticking probability $\omega_{\rm d}^0$ as
\begin{equation}
\omega_d^{\rm eff}{\rm (th)}=\omega_d^{11} (1- R),
\end{equation}
where $R$ is the muon reactivation coefficient expressing the probability that 
the muon is shaken off during the $(^3{\rm He}\mu)$ atom finally comes to rest.
Ref.~\cite{Balin2011} summarized the theoretical value of $R$ as 
$R=0.10 \pm 0.01 (\varphi=0.07)$ referring to the 
work~\cite{Markushin1988,Struensee1988,Takahashi1988}.
Therefore, our $\omega_{\rm d}^{\rm eff}{\rm (th)}$ agrees with the
observed $\omega_{\rm d}^{\rm eff}{\rm (exp)}$ barely within the quoted errors. 

\begin{table}
\caption{The sticking probabilities $\omega_{\rm d}^{Jv}$ for the
$(Jv)=(11)$ and $(10)$ states of the $dd\mu$ molecule. }
\label{tab:vdtanVI}
  \begin{tabular}{ccccc}
\noalign{\vskip 0.1 true cm}
    \hline
    \hline
\noalign{\vskip 0.1 true cm}
$p$-wave $S(E)$ factor & $\quad$ &  $\;\;\omega_{\rm d}^{11}\;$ &  
   $\quad\;$ & $\;\omega_{\rm d}^{10}\;$   \\
\noalign{\vskip 0.1 true cm}
\noalign{\vskip 0.0 true cm}
    \hline
\noalign{\vskip 0.05true cm}
\noalign{\vskip 0.05true cm}
 Angulo+ 1998  & &\!\!\!\! 0.133(1)  & &0.133(1)    \\
\noalign{\vskip 0.1 true cm}
 Nebia+ 2002  & &\!\!\!\! 0.133(1)  & &0.133(1)    \\
\noalign{\vskip 0.1 true cm}
 Arai+ 2011  & &\!\!\!\! 0.133(1)  & &\!\!\!\! 0.133(1)    \\
\noalign{\vskip 0.1 true cm}
 Tumino+ 2014  & &\!\!\!\! 0.133(1)  & &\!\!\!\! 0.133(1)    \\
\noalign{\vskip 0.1 true cm}
 Solovyev 2024  & &\!\!\!\! 0.133(1)  & &\!\!\!\! 0.133(1)    \\
\noalign{\vskip 0.1 true cm}
\hline
\hline
\noalign{\vskip 0.1 true cm}
\end{tabular}
\end{table}

\begin{table}
\caption{ 
The individual sticking probability $\omega_{\mathrm{d}}^{Jv}(n l)$ 
to the $(^3{\rm He}\mu)_{nl}$ states from 
the $(Jv)=(11)$ and (10) states of the $dd\mu$ molecule,
in the cases of the $S(E)$ factors of Tumino+ 2014.
They are the average of the results using the 20 sets of the nuclear potentials
in Table~\ref{tab:vdtan} with relative deviations less than 2 \%.
The numbers in the parentheses are given by 
Bogdanova {\it et al.}~\cite{Bogdanova1985}.
}
\label{tab:rnl}
  \begin{tabular}{cccccc}
\noalign{\vskip 0.1 true cm}
    \hline
    \hline
\noalign{\vskip 0.1 true cm}
$ \quad  nl \quad \quad \quad$ &  
$\quad \omega_{d}^{11}(nl) $ &     & \quad \quad \; & 
$ \quad  \omega_{d}^{10}(nl) $&      \\
\noalign{\vskip 0.1 true cm}
    \hline
\noalign{\vskip 0.1 true cm}
   1s     & 0.0942 & (0.0947) & \quad  & 0.0941 & (0.0936)     \\
\noalign{\vskip 0.05 true cm}
   2s     & 0.0126 & (0.0126) & \quad  & 0.0126 & (0.0125)     \\
\noalign{\vskip 0.05 true cm}
   2p     & 0.0102 & (0.0101)& \quad  & 0.0103 & (0.0100)     \\
\noalign{\vskip 0.05 true cm}
   3s     & 0.0037 & (0.0037) & \quad  & 0.0037 & (0.0037)     \\
   \noalign{\vskip 0.05 true cm}
   3p     & 0.0037 & (0.0036) & \quad  & 0.0036 & (0.0035)     \\
   \noalign{\vskip 0.05 true cm}
   3d     & 0.0003 & (0.0003) & \quad  & 0.0003 & (0.0003)     \\
   \noalign{\vskip 0.05 true cm}
   4s     & 0.0016 & (0.0016) & \quad  & 0.0016 & (0.0015)     \\
   \noalign{\vskip 0.05 true cm}
   4p     & 0.0016 & (0.0015) & \quad  & 0.0016 & (0.0015)     \\
   \noalign{\vskip 0.05 true cm}
   4d+4f  & 0.0002 & (0.0002) & \quad  & 0.0002 & (0.0002)     \\
   \noalign{\vskip 0.05 true cm}
   $n\geq5$     & 0.0050 & (0.0052) & \quad  & 0.0050 & (0.0051)     \\
   \noalign{\vskip 0.05 true cm}
\hline
\noalign{\vskip 0.1true cm}
   total     & 0.1330 & \!(0.133)  & \quad  & 0.1330 &\! (0.132)      \\
\noalign{\vskip 0.05 true cm}
\hline
\hline
\noalign{\vskip 0.1 true cm}
\end{tabular}
\end{table}

The  sticking probability to each $(^3{\rm He}\mu)_{nl}$ state,
say $\omega_{\rm d}^{Jv}(nl)$, is given by replacing 
$\lambda_{Jv}^{(^3{\rm He}n\mu)}(\rm bound)$ at the numerator in Eq.~(5.1)
with  $r_{Jv,nl}^{(c=5)}$ in Eq.~(4.10). 
Table VIII contains the $\omega_{\rm d}^{11}(nl)$ and 
$\omega_{\rm d}^{10}(nl)$ calculated with the
potential set A1 in the case of Tumino+ 2014, while
the numbers in the parentheses are given by 
Bogdanova {\it et al.}~\cite{Bogdanova1985}; obviously, close to each other.

%
\section{Fusion rate of \boldmath{\lowercase{$dd\mu$}} molecule (\lowercase{iii}): \\
\vskip 0.05cm
     \boldmath{$T$}-matrix model on channels $\lowercase{c=}$ 4 and 7}

In this section, we calculate the fusion rates of the $(dd\mu)_{J=1,v}$ 
molecule employing {\it method} iii), namely, using the tractable three-body 
$T$-matrix model~\cite{Wu2024}
taking channels $c=4$ and 7 (Fig.~2) for the description of the outgoing particles.
One reason is we shall calculate the momentum and energy spectra of the emitted 
muon in the next section.  
Note that the muon is emitted, along the coordinates ${\bf R}_4$ and ${\bf R}_7$ 
in Fig.~2, from the c.m. of the $dd\mu$ molecule that is finally almost at rest 
in the laboratory system before fusion. 

We consider the following reactions $(i=1 - N)$,
\begin{eqnarray}
\!\!\!  (dd\mu)_{Jv} \to (^3{\rm He}n)_{il} + \mu + 4.03 \:\mbox{MeV},
\end{eqnarray}
\begin{eqnarray}
\!\!\! (dd\mu)_{Jv} \to (tp)_{il} + \mu + 3.27 \:\mbox{MeV}, 
\end{eqnarray}
where $(^3{\rm He}n)_{il}$ and $(tp)_{il}$ denote the 
$^3{\rm He}$-$n$ and $t$-$p$ continuum-discretized states along ${\bf r}_4$ and ${\bf r}_7$, respectively.
Note that there is no bound state with $l \ge 1$.  

In the study of the $\mu$CF of $(dt\mu)$ molecule in Ref.~\cite{Wu2024}, we experienced the above type of reactions using the $T$-matrix model.
Similarly to the study, we discretize the $^3{\rm He}$-$n$ continuum into $N=200$ bins and correspondingly for $^3{\rm He}n$-$\mu$, keeping the energy conservation (cf. Fig.~13 of Ref.~\cite{Kamimura2023}).

To formulate the fusion rate and the $T$ matrix of the reaction (6.1), we modify Eqs. (4.6) and (4.7) for channel $c=5$ to $c=4$ and generate the following expression, 
with the use of similar notations,\footnote{
We take the plane wave for the relative motion between $(^3{\rm He}n)_{il}$ and $\mu$. The reason why it is not necessary to employ the Coulombic wave function is explained in Appendix of Ref.~\cite{Kamimura2023} in the case of $(^4{\rm He}n)_{il}$ and $\mu$.} 
\begin{eqnarray}
&&\!\!\!\!\!\!\!\!
{\widetilde T}^{(c=4)}_{Jv,ilm}({\widetilde {\bf K}}_i) = 
    \langle e^{ i {\widetilde {\bf K}}_i \cdot {\bf R}_4 } \,
      {\widetilde \phi}_{ilm}({\bf r}_4)
    |\,  {\cal V}^{\rm (cp)}_{\!^3{\rm He} n, dd} \,| \,
 \Phi^{\rm (opt)}_{JM, v}(dd\mu)  \,\rangle,   \nonumber \\ \\
 && \!\!\!\!\!\!\!\!
{\widetilde r}_{Jv,il}^{\,(c=4)} ={v}^{(4)}_{il}
  \!  \left( \frac{\mu_{R_4}}{2 \pi \hbar^2} \right)^{\!2} 
     \! \big| S^{\rm (cp)}_1 \, \big|^2
   \sum _{m}\! \int \big| {T}^{(c=4)}_{Jv,ilm}
   ({\widetilde {\bf K}}_i) \,\big|^2   {\rm d}{\widehat {\widetilde {\bf K}}_i},
    \nonumber \\  
\label{eq:T4-cont1}
\end{eqnarray}
and similarly for the $t$-$p$ channel of $c=7$.
 
In Eq.~(6.3), the energy of the plane wave 
$e^{ i {\widetilde {\bf K}}_i \cdot {\bf R}_4 }$ and that of the 
$^3{\rm He}$-$n$ relative motion ${\widetilde \phi}_{ilm}({\bf r}_4)\,(i=1-N)$
should satisfy the energy conservation (cf. Eq.~(4.8)),
\begin{eqnarray}
 \hbar^2{\widetilde K}_i^2/2\mu_{R_4}  + {\widetilde \varepsilon}_{i}
 = E_{Jv}^{\rm (real)} + 4.03 \, \mbox{MeV}. 
\label{eq:E-conservation-2}
\end{eqnarray} 
%

In the same way as used in the $^4{\rm He}n\mu$ system in Ref.~\cite{Wu2024}, we discretize the momentum ${\widetilde  K}$-space for the relative \mbox{($^3{\rm He}n)$-$\mu$} motion into $N=200$ bins between $\hbar {\widetilde K}_0=0$ and $\hbar {\widetilde K}_N=6 \,{\rm MeV}/c\, ({\widetilde E}_N=175\, {\rm keV})$, with a constant bin size $\Delta {\widetilde K}=6/200\, {\rm MeV}/c$.
This is sufficiently precise for deriving the muon spectrum with a smooth function, especially in the peak energy region.
Correspondingly, the momentum $k$-space for the relative $^3{\rm He}$-$n$ motion, energetically having 175 keV-width below the $Q$-value (4.03 MeV), is discretized into $N=200$ bins under the energy conservation Eq.~(6.5), but with unequal bin sizes (cf. Fig.~8 in Ref.~\cite{Wu2024}).

Similarly, we obtain the following expression for the reaction (6.2) by modifying Eqs.~(6.3) and (6.4) to the case of channel $c=7$, 
\begin{eqnarray}
&&\!\!\!\!\!\!\!\!
{\widetilde T}^{(c=7)}_{Jv,ilm}({\widetilde {\bf K}}_i) = 
    \langle e^{ i {\widetilde {\bf K}}_i \cdot {\bf R}_7 } \,
      {\widetilde \phi}_{ilm}({\bf r}_7)
    |\,  {\cal V}^{\rm (cp)}_{tp, dd} \,| \,
 \Phi^{\rm (opt)}_{JM, v}(dd\mu)  \,\rangle,   \nonumber \\ \\
 && \!\!\!\!\!\!\!\!
{\widetilde r}_{Jv,il}^{\,(c=7)} ={v}^{(7)}_{il}
  \!  \left( \frac{\mu_{R_7}}{2 \pi \hbar^2} \right)^{\!2} 
     \! \big| S^{\rm (cp)}_1 \, \big|^2
   \sum _{m}\! \int \big| {T}^{(c=7)}_{Jv,ilm}
   ({\widetilde {\bf K}}_i) \,\big|^2   {\rm d}{\widehat {\widetilde {\bf K}}_i}.
    \nonumber \\  
\label{eq:T7-cont1}
\end{eqnarray}
When calculating the $T$-matrix elements (6.3) and (6.6),
the method of Eqs.~(4.16) and (4.17) is useful.

\begin{table*}
\caption{
Fusion rates
$\lambda_{Jv}^{(^3{\rm He}n\mu)}$, $\lambda_{Jv}^{(tp\mu)}$
and their sum $\lambda_{Jv}$ 
of the $(dd\mu)_{J,v}$ states with $J=1, v=1$ and $0$, 
calculated on the channels $c=4$ and $7$ using the 20 potential sets $A1$ 
to $E4$ (cf. Table~\ref{tab:vdtan}).
All in unit of $10^8 {\rm s}^{-1}$.
}
\label{tab:vdtanVIII}
  \begin{tabular}{cccccccc}
\noalign{\vskip 0.1 true cm}
    \hline
    \hline
\noalign{\vskip 0.1 true cm}
$p$-wave   &  $c=4$  & $c=7$  & $c=4\& 7$   & & $c=4$ &  $c=7$
&  $c=4\& 7$   \\
\noalign{\vskip 0.02true cm}
\;\;\;\;$S(E)$ factor   \;\;\;\;   &  $ \;\; \lambda_{11}^{(^3{\rm He}n\mu)}$  \;\; 
& $\lambda_{11}^{(tp\mu)}$  \;\;  &  \;\; 
$\lambda_{11}$  \;\; & $   $  \;\; &  \;\; 
$\lambda_{10}^{(^3{\rm He}n\mu)}$  \;\; &  \;\; 
$\lambda_{10}^{(tp\mu)}$  \;\; &  \;\; $\lambda_{10}$  \;\; \\
\noalign{\vskip 0.1 true cm}
\noalign{\vskip 0.0 true cm}
    \hline
\noalign{\vskip 0.1 true cm}
\noalign{\vskip 0.05true cm}
Angulo+ 1998 &  0.84(3) \: &  0.94(3)\:\,  & \;\;1.8(1) \:\,& & 
             2.6(3) \: & 2.9(3) \: & \: 5.6(6) \:  \\
\noalign{\vskip 0.05true cm}
Nebia+ 2002 &  2.5(1) \: &  1.7(1)\:\,  & \;\;4.2(1) \:\,& & 
             7.7(1) \: & 5.2(1) \: & \: 13.0(2) \:  \\
\noalign{\vskip 0.05true cm}
Arai+ 2011  &  2.9(1) \: &  2.1(1)\:\,  & \;\; 5.0(1) \:\,& & 
             9.0(1) \: & 6.7(1) \: &  15.6(1) \:  \\
\noalign{\vskip 0.1 true cm}
Tumino+ 2014 &  2.1(1) \: &  2.1(1)\:\,  & \;\;4.2(1) \:\,& & 
             6.5(3) \: & 6.4(3) \: &  13.1(5) \:  \\
\noalign{\vskip 0.05true cm}
Solovyev 2024   &  1.5(1) \: &  1.2(1)\:\,  & \;\;2.7(1) \:\,& & 
             4.7(1) \: & 3.6(1) \: & \: 8.3(2) \:  \\
\noalign{\vskip 0.1 true cm}
\hline
\hline
\noalign{\vskip 0.1 true cm}
\end{tabular}
\end{table*}

The sum of the transition rates,
\begin{eqnarray}
\quad \lambda_{J=1,v}^{({\rm ^3He}n\mu)}&=& 
 \sum_{il} {\widetilde r}^{\,(c=4)}_{J=1,v,\,il}\, , \quad (c=4),  \\
\quad \lambda_{J=1,v}^{(tp\mu)}&=& 
 \sum_{il} {\widetilde r}^{\,(c=7)}_{J=1,v,\,il}\, , \quad (c=7),  \\
\quad \lambda_{J=1,v}&=& 
\quad \lambda_{J=1,v}^{({\rm ^3He}n\mu)} + \lambda_{J=1,v}^{(tp\mu)}
\label{eq:fusion-rate-muon-47}
\end{eqnarray}
give the fusion rates of the $(dd\mu)_{J=1,v}$ molecule, respectively. 
The contributions from the states with $l \neq 1$  are negligible.

Table \ref{tab:vdtanVIII} lists the fusion rates $\lambda_{Jv}^{({\rm ^3He}\mu)}$, 
$\lambda_{Jv}^{(tp\mu)}$, and their sum $\lambda_{Jv}$ for the states with $J=1, v=0$ 
and 1, using the 20  potential sets (cf. Table~\ref{tab:vdtan}).
We see that $\lambda_{Jv}^{({\rm ^3He}\mu)}$ and $\lambda_{Jv}^{(tp\mu)}$ agree well 
with those in Table \ref{tab:vdtanV}, as long as the comparisons are conducted separately 
with those of Anglo+ 1998, Nebia+ 2002,  Arai+ 2011, Tumino+ 2014, and Solovyev 2024.
\mbox{Similarly,} $\lambda_{J,v}$ agree with $\lambda_{Jv}^{({\rm opt})}$ 
in \mbox{Table~\ref{tab:lambda-opm}.}
These agreements indicate the validity of the three methods for calculating 
the \mbox{fusion} rate of the reactions (1.3) and (1.4).

We therefore summarize, as in \mbox{Table~\ref{tab:lambda-muon-spec}},
those fusion rates $\lambda_{11}$ calculted with the five types of 
$p$-wave $S(E)$ factors, together with the effective
fusion rates $\widetilde{\lambda}_{\rm f}$ estimated by 
Eq.~(\ref{eq:effective-fusion}). 
We see a significant difference in those fusion rates 
between the five cases of $S(E)$ factors employed; 
the effective fusion rates $\widetilde{\lambda}_{\rm f}$ 
are as widely distributed as $(2.0 - 5.3)\times 10^8 {\rm s}^{-1}$ 
although they cover the observed values in Table~III.
More precise experimental determination is strongly
required of the $p$-wave $S(E)$ factor for the reactions (1.1) and (1.2) 
at low energies.

\begin{table}[h]
\centering
\caption{
Summary of calculated fusion rate $\lambda_{11}$ and the effective fusion rate
$\widetilde{\lambda}_{\rm f}$ in the present work (cf. Tables IV, VI and IX).
They are to be compared with the
\mbox{experimental} data in Table \ref{tab:lambda-ref}.
Same meaning for the numbers with \mbox{superscript(*)} as in Table \ref{tab:lambda-ref}.
All the rates are in units of $10^8{\rm s}^{-1}$.  
}
\begin{tabular}{cclcl}
       \noalign{\vskip 0.1 true cm}
       \hline
       \hline
\noalign{\vskip 0.1 true cm}
 $p$-wave  $S(E)$ factor    & \qquad \qquad & $\lambda_{11}\quad$   
        &\qquad \qquad  & $\quad \widetilde{\lambda}_{\rm f} \qquad$  \\
\noalign{\vskip 0.1 true cm}
\hline
\noalign{\vskip 0.1 true cm}
 Angulo+ 1998  & &\!\!\!\!  1.8(1)  & & $\;\;\;2.0^*$    \\
\noalign{\vskip 0.1 true cm}
 Nebia+ 2002  & &\!\!\!\! 4.2(1)  & & $\;\;\;4.4^*$    \\
\noalign{\vskip 0.1 true cm}
 Arai+ 2011  & &\!\!\!\! 5.1(1)  & & $\;\;\;5.3^*$    \\
\noalign{\vskip 0.1 true cm}
 Tumino+ 2014  & &\!\!\!\! 4.2(1)  & & $\;\;\;4.4^*$    \\
\noalign{\vskip 0.1 true cm}
 Solovyev 2024  & &\!\!\!\! 2.7(1)  & & $\;\;\;2.9^*$    \\
\noalign{\vskip 0.1 true cm}
\noalign{\vskip 0.05 true cm}
\hline
\hline
\end{tabular}
\label{tab:lambda-muon-spec}
\end{table}


\section{Momentum and energy spectra of emitted muons}

This section presents the momentum and energy spectra
of the muons emitted in reactions (6.1) and (6.2).
The momentum spectrum,  $r_{Jv}(K)$, 
is obtained by smoothing 
\mbox{${\widetilde r}^{\,(c=4)}_{Jv,\,il}+{\widetilde r}^{\,(c=7)}_{Jv,\,il}$}
in Eqs. (6.8) and (6.9) as,
\begin{eqnarray}
\lambda_{Jv} &\!\!=\!\!& 
 \sum_{il} \Big( \frac{ {\widetilde r}^{\,(c=4)}_{Jv,\,il}
                       +{\widetilde r}^{\,(c=7)}_{Jv,\,il} }{\Delta K} \Big) \,  
                        \mathit{\Delta}K              \nonumber \\
&& \!\!\!\!\!\!\!\!\! \stackrel{\mathit{\Delta}\!K \to 0}{\longrightarrow}
  \int_0^{K_N} \! {r}_{Jv}(K)\, {\rm d} K ,   
\label{eq:fusion-rate-muon-47}
\end{eqnarray}
\noindent
where the present case $\mathit{\Delta} K=0.03$ MeV/$c$ is sufficiently small.
The energy distribution, ${\bar r}(E)$, is derived as 
\begin{eqnarray}
{\bar r}_{Jv}(E) \,{\rm d}E =r_{Jv}(K)\, {\rm d} K,\quad
E=\hbar^2 K^2/2\mu_{{\rm R}_4}.
\label{eq:muon-E-spec-1}
\end{eqnarray}

\begin{figure}
\vskip -0.1cm
\centering
\includegraphics[width=0.53\textwidth]{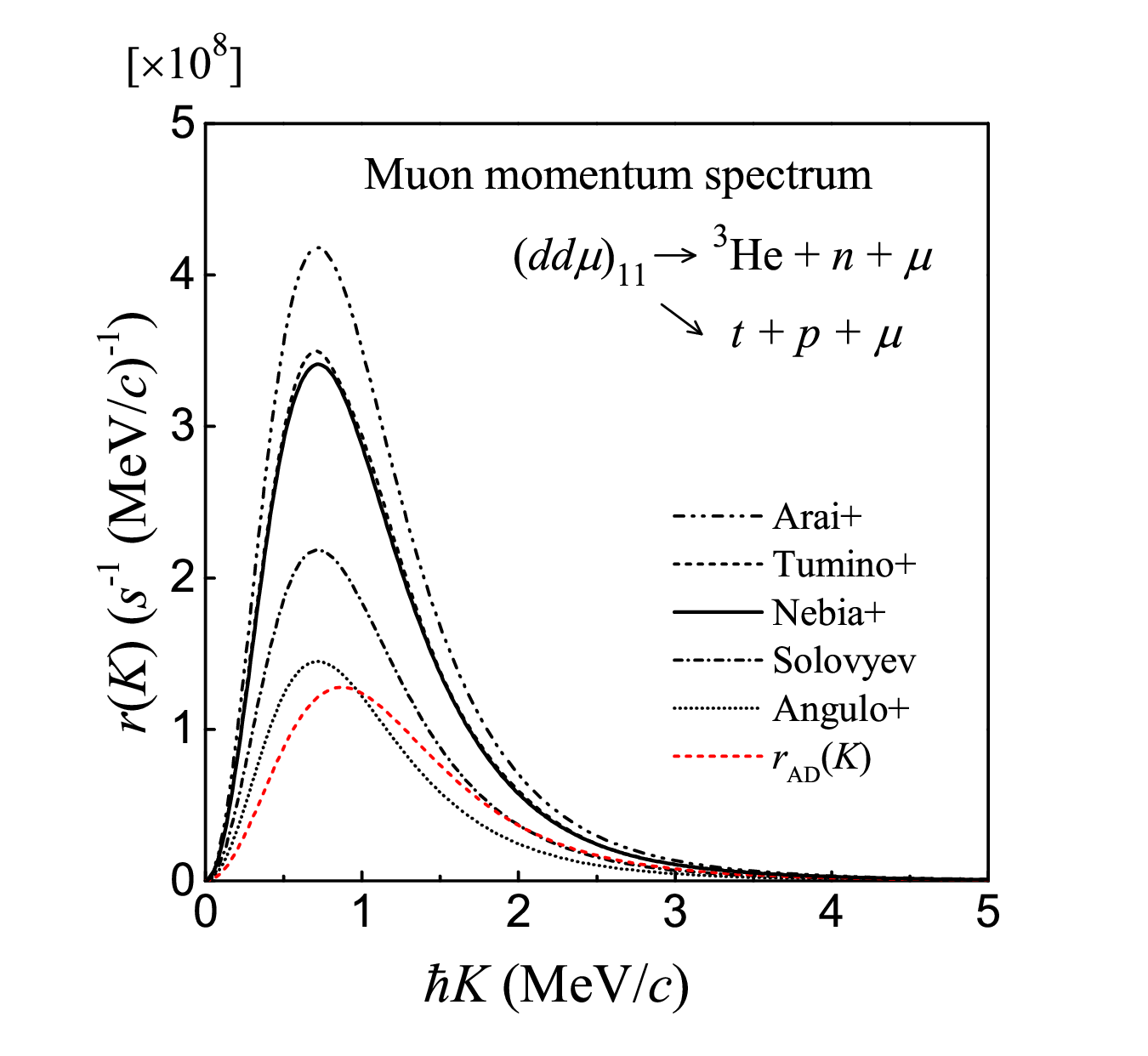}
\vskip -0.1cm
\caption{
Momentum spectrum $r_{Jv}(K)$ in Eq.~(7.1) of muons emitted from the $J=v=1$ state, 
calculated using the potential set A1 in Table~\ref{tab:vdtan}.
The lines for the $J=1, v=0$ state have almost the same shape, 
but the magnitudes are nearly 3.1 times larger. 
The dotted red line shows the adiabatic limit Eq. (7.3):
the magnitude is normalized to Angulo+ for comparison m to have the same
$\hbar K$-integrated value.
}
\label{fig:rK}
\end{figure}

\begin{figure}
\vskip -0.1cm
\centering
\includegraphics[width=0.53\textwidth]{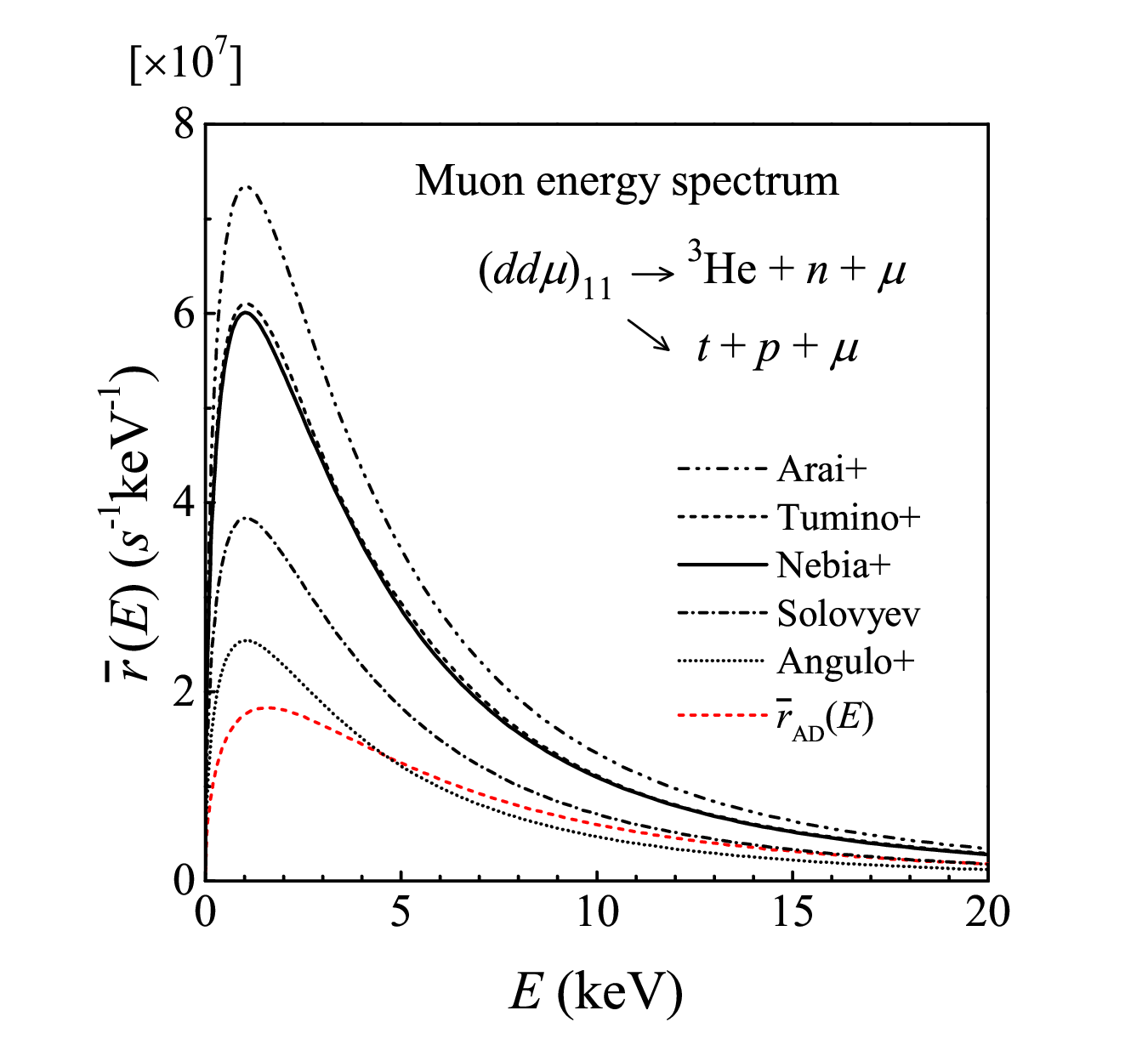}
\vskip -0.1cm
\caption{
Energy spectrum ${\bar r}_{Jv}(E)$ in Eq. (7.2) 
of muons emitted from the $J=v=1$ state, 
calculated using the potential set A1 in Table~\ref{tab:vdtan}.
The lines for the $J=1, v=0$ state have almost the same shape, 
but the magnitudes are nearly 3.1 times larger. 
The dotted red line shows the adiabatic limit Eq. (7.4):
the magnitude is normalized to Angulo+ to have the same
$E$-integrated value.
}
\label{fig:rE}
\end{figure}

Figs.~\ref{fig:rK} and~\ref{fig:rE} illustrate the muon momentum spectrum 
$r_{Jv}(K)$ and the energy spectrum ${\bar r}_{Jv}(E)$
of the $J=v=1$ state, calculated using the nuclear potentials 
which reproduce the five cases of the $p$-wave $S(E)$ factors 
individually (cf. Figs.~4 and 5).
Here, we use potential set A1 in Table~\ref{tab:vdtan} for each
case of the $S(E)$ factor, whereas the lines using other sets give very similar 
results but are omitted to avoid complexity.

The function forms of the five lines in Fig.~\ref{fig:rK} (Fig.~\ref{fig:rE})
are almost the same to each other 
since the $p$-wave $S(E)$ factors in Fig.~1 have nearly the same shape 
at lower energies. 
Note that, in the figures, the $K\hbar$-integrated ($E$-integrated) value of 
each line is just the fusion rate $\lambda_{11}$, and  therefore
height of the line is  proportional to $\lambda_{11}$.  

The dotted red lines in Figs.~\ref{fig:rK} and~\ref{fig:rE}
show the muon momentum and energy spectra when taking the adiabatic approximation 
for the \mbox{$d$-$d$} relative motion just before  the fusion reaction 
and the sudden approximation after the fusion process.
The wave function of the \mbox{$(dd)$-$\mu$} relative motion
is simply given by $\propto e^{-R_4/a_0}$ with $a_0=131$ fm
as that of the $({\rm He}\mu)_{1s}$ atom,
which has the mean kinetic energy of 10.9 keV.
In the adiabatic approximation, the momentum spectrum of 
emitted muon, namely the reaction rate $r_{\rm AD}(K)$,
is assumed to have the same function form 
of the muon momentum distribution of the $({\rm He}\mu)_{1s}$ atom,
\begin{eqnarray}
r_{\rm AD}(K)  \propto {K^2}/{(1+ K^2 a^2)^4}. \quad 
\label{eq:rNK-simulate} 
\end{eqnarray}
The energy spectrum, ${\bar r}_{\rm AD}(E)$, is given by Eq.~(7.2) as
\begin{eqnarray}
{\bar r}_{\rm AD}(E) \, \propto\, {K}/{(1+ K^2 a^2)^4}.  \qquad
\end{eqnarray}

Here, the magnitude of $r_{\rm AD}(K)$ (${\bar r}_{\rm AD}(E))$ is normalized 
to Angulo+ for comparison to have the same $\hbar K$-integrated ($E$-integrated) value.
We note that, in both Figs.~9 and 10, the lines for \mbox{Anglo+} are  
significantly shifted to the left from the dotted red lines, 
with the peak heights much enhanced. This indicates that, 
in the actual fusion time, the muon is spatially  much less attracted 
by the $d$-$d$ system, which is moving in a much more `wider' region than 
that of the adiabatic case.

In Table~\ref{tab:av-f4}, the peak and average energies
of the muon energy spectra ${\bar r}_{J=v=1}(E)$ in Fig.~10 
are listed (the values are the same for $J=1,v=0$ state).
It is to be emphasized that the peak energy is located at 1.0 keV,
much smaller than the average energy of 8.2 keV which is caused by 
the long high-energy tail seen in Fig.~\ref{fig:rE}.

Observation of the emitted  muon spectra can provide rich information 
on the few-body quantum mechanics of the fusion processes. 
Fundamental experiments are in progress
by Refs.~\cite{Strasser1993,Strasser2000,Yamashita2021,Okutsu2021} with the use of 
a two-layer solid hydrogen film target from which \mbox{a half} of the released 
muons immediately go into the free space.
The  muon's spectrum is calculated for the first time, 
and will be helpful for  future experiments that generates an ultra-slow 
negative muon beam using the \mbox{$d$-$d$} $\mu$CF for various applications.


\begin{table}
\caption{Property of the muon energy spectrum ${\bar r}_{J=v=1}(E)$ 
with the use of the potential set A1 in Table~\ref{tab:vdtan}. 
Use of the other the potential sets gives almost the same results.
The last line is for the adiabatic limit given in Eq.~(7.4).
}
\begin{center}
\begin{tabular}{ccccccc}
\noalign{\vskip -0.1 true cm}
\hline \hline
\noalign{\vskip 0.1 true cm}
  $p$-wave & $ \;$ &  Peak  &   &
Average    &  & Peak   \\
\noalign{\vskip 0.05true cm}
$S(E)$ factor  &  &   energy &   &  energy   &  &  strength  \\
\noalign{\vskip 0.01true cm}
  &  &  (keV) & $\;$  &  (keV)   &  &
    $({\rm s}\cdot{\rm keV})^{-1}$  \\
\noalign{\vskip 0.1true cm}
\hline
\noalign{\vskip 0.1true cm}
Angulo+ 1998 &  & 1.0    &   &  8.2  &  & $ 2.5 \times 10^{7}$ \\
\noalign{\vskip 0.1true cm}
Nebia+ 2002 &   & 1.0    &   &  8.2  &  & $ 6.0 \times 10^{7}$ \\
\noalign{\vskip 0.1true cm}
Arai+ 2011  &  &  1.0 &   &  8.2   &  & $  7.3 \times 10^{7}$ \\
\noalign{\vskip 0.1true cm}
Tumino+ 2014 &  & 1.0  &   &  8.2  &  & $  6.1 \times 10^{7}$ \\
\noalign{\vskip 0.1true cm}
Solovyev 2024 & & 1.0    &   & 8.2 &  & $ 3.8 \times 10^{7}$ \\
\noalign{\vskip 0.1true cm}
Adiabatic &  & 1.6 &   & 10.9   &  &  \\
\noalign{\vskip 0.1 true cm}
\hline
\hline
\noalign{\vskip -0.3 true cm}
\end{tabular}
\label{tab:av-f4}
\end{center}
\end{table}

It is to be noted here that the muon-sticking to $^3{\rm He}$ gives little effect 
in the important energy (momentum) region in Figs.~\ref{fig:rK} and~\ref{fig:rE}.
The reason is as follows: As discussed in Sec.~IV below Eq.~(4.25), 
the $^3{\rm He}$ particles escape from the $1s$-like muon cloud
after the fusion with a speed $v_{^3{\rm He}}/c=0.024$,  and therefore
the muons with nearly the same speed have the probability of sticking to
$^3{\rm He}$. The corresponding energy of the muons is 
$\approx \! 30$ keV and the momentum is $\approx 2.5 \,{\rm MeV}/c$,
which is much higher than the peak region.

\section{Violation of  charge symmetry in  
{\boldmath $\lowercase{p\,}$}-wave 
 \mbox{\boldmath {$\lowercase{d+d}$}} and 
\mbox{\boldmath {$\lowercase{d+d+\mu}$}}
reactions}  

As mentioned in the Introduction, the ratio $R_S$ (1.5) of the $p$-wave $S(E)$ factors at $E \to 0$ has been used historically 
in the studies of the violation of the charge symmetry in the reactions (1.2) and (1.3) with  $R_S \simeq 1.4$ shown.
The origin of this large value of $R_S$ was explained by Hale~\cite{Hale1990}, using the $R$-matrix analysis of the $A=4$ system,  as the result of  the isospin mixing between the broad $J=1^-$ levels at $E_{\rm x}=23.64$ MeV $(T=0)$ and 24.25 MeV $(T=1)$ being located near the $d$+$d$ threshold.

Interestingly, Bogdanova {\it et al.}~\cite{Bogdanova1982}
showed that the ratio $R_S$ is equal to the ratio $R_Y$ in Eq.~(1.6) under the factorization approximation of the $dd\mu$ fusion rate (cf. their Eq.~(4)).
Actually, Balin {\it et al.}~\cite{Balin1984} obtained
$R_Y=1.39 \pm 0.04$ in the $dd\mu$ fusion experiment.

\begin{figure}
\setlength{\abovecaptionskip}{0.cm}
\setlength{\belowcaptionskip}{-0.cm}
\centering
\includegraphics[width=0.40\textwidth]{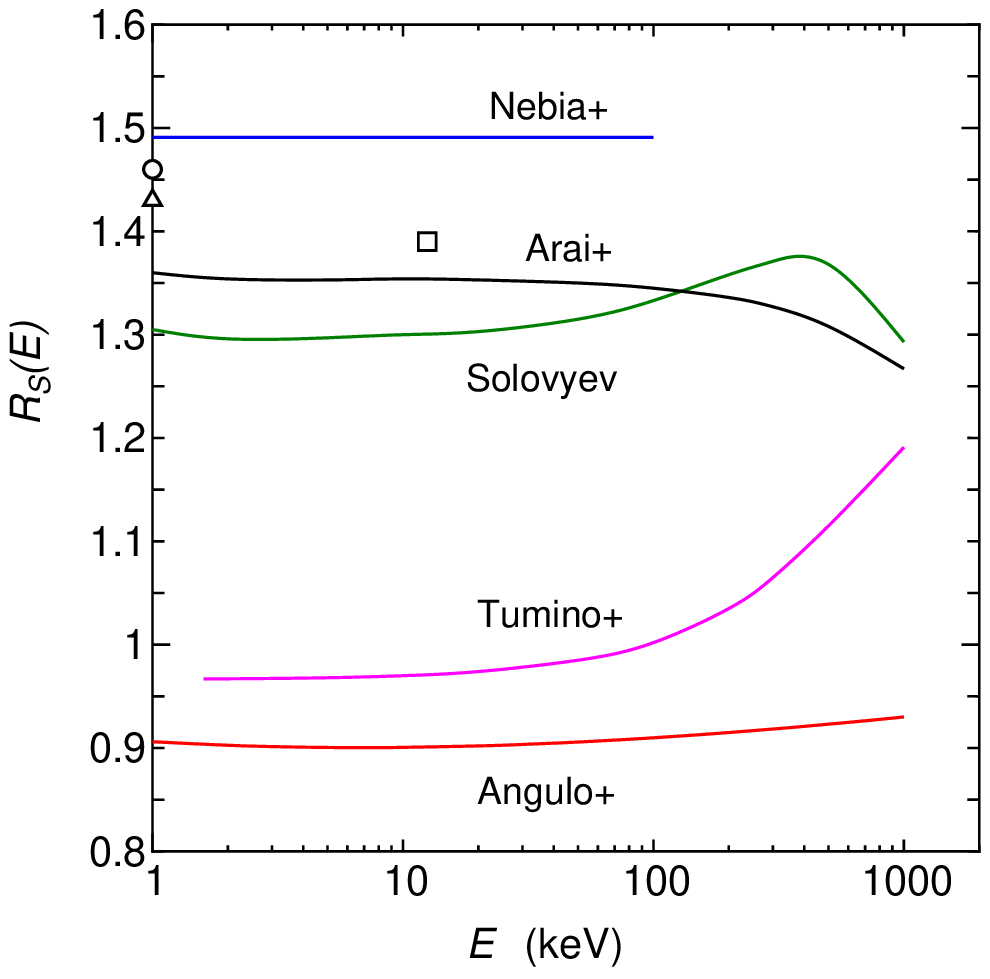}
\caption{Energy dependence of the $S$-factor ratio $R_S(E)$ of Eq.~(8.1),
with respect to Angulo+ 1998, Nebia+ 2002, Arai+ 2011, Tumino+ 2014,
and Solovyev 2024. The black circle ($R_S(E)=1.46$), triangle (1.39), 
and box (1.39) are respectively by Refs.~\cite{Adya1981,Hale1990} at 
$E \to 0$ and Ref.~\cite{Fletcher1994} at $E=12.5$ keV.
}
\label{fig:violation}
\end{figure}

Now, we know the $p$-wave $S(E)$ factors for $E=$ 1 keV to 1 MeV 
given by five experimental and \mbox{theoretical} 
studies~\cite{Angulo1998,Nebia2002,Arai2011,Tumino2014,Solovyev2024}
as illustrated in Fig.~\ref{fig:8line-sfactor}.
Using those $S(E)$ factors, say $S_{^3{\rm He}+n}(E)$ and $S_{t+p}(E)$, 
we introduce the energy-dependent ratio $R_S(E)$ as  
\begin{eqnarray}
  &&  R_S(E) = S_{^3{\rm He}+n}(E)/S_{t+p}(E),
\end{eqnarray}
which are illustrated in Fig.~\ref{fig:violation} together with 
three $R_S$ values by Refs.~\cite{Adya1981,Hale1990,Fletcher1994}.
It is noticeable that, in the region of $E$ up to 100 keV,
these five lines are almost constant, and divided into two groups of 
\mbox{$R_S \simeq 1.3$ - $1.5$} with the large charge symmetry violation 
and of $R_S \simeq$ \mbox{$0.9$ - $1.0$}.

%

Employing these $p$-wave $S(E)$ factors, we calculate the fusion rates of 
the $J=v=1$ states, $\lambda_{11}^{(^3{\rm He}n\mu)}$  and   
$\lambda_{11}^{(tp\mu)}$, as shown 
in Table~\ref{tab:vdtanV} and ~\ref{tab:vdtanVIII}, which give the ratio $R_Y$ as
\begin{eqnarray}
  &&  R_Y = \lambda_{11}^{(^3{\rm He}n\mu)}/\lambda_{11}^{(tp\mu)},
\end{eqnarray}
after taking the average over the 20 sets of the interaction
parameters (cf. Table~\ref{tab:vdtan}) and the calculation channels  
$c=5\&8$  and $c=4\&7$.
Table~\ref{tab:ratio-RS-RY} summarizes the results\footnote{
The reason of the small deviation of $R_Y$ is because
the average of the `relative ratio' is taken (cf. the case of sticking probabilities 
in Table \ref{tab:vdtanVI}).}.
It is remarkable to see $R_S( E \!=\! 1 \, {\rm keV)}\!=\!R_Y$, which supports 
the property of  $R_S=R_Y$ argued by Bogdanova {\it et al.}~\cite{Bogdanova1982} 
under the factorization approximation of  the $dd\mu$ fusion rate.

\begin{table}
\caption{$R_S( E \!=\! 1 \, {\rm keV)}$ defined by Eq.~(8.1) and $R_Y$ by Eq.~(1.6).
The latter is calculated using the $p$-wave $S(E)$ factors of
Angulo+ 1998, Nebia+ 2002, Arai+ 2011, Tumino+ 2014, and Solovyev 2024.
$R_Y=1.455 (11)$ is the latest observed value by
Balin {\it et al.}~\cite{Balin2011}.
}
\begin{center}
\begin{tabular}{ccccc}
\noalign{\vskip -0.2true cm}
\hline \hline
\noalign{\vskip 0.3 true cm}
$p$-wave $S(E)$ factor  & \quad\qquad &  $R_S(E=1\,{\rm  keV})$   &   
& \qquad  $R_Y$ \quad  \quad \\
\noalign{\vskip 0.3true cm}
\hline
\noalign{\vskip 0.1true cm}
Angulo+ 1998 &  &  0.908    &   & \qquad  $0.91 \pm 0.03$ \quad \quad    \\
\noalign{\vskip 0.1true cm}
Nebia+ 2002 &   & 1.491     &   & \qquad  $1.49 \pm 0.02$ \quad \quad   \\
\noalign{\vskip 0.1true cm}
Arai+ 2011  &  &  1.360   &   &  \qquad $1.36 \pm 0.02$   \quad \quad    \\
\noalign{\vskip 0.1true cm}
Tumino+ 2014 &  & \: 0.967\footnote{
at $E=$ 1.6 keV}    &   &\qquad   $1.03 \pm 0.05$  \quad \quad    \\
\noalign{\vskip 0.1true cm}
Solovyev 2024 & & 1.305      &   &\qquad  $1.32 \pm 0.02$\quad \quad    \\
\noalign{\vskip 0.1 true cm}
\hline
\hline
\noalign{\vskip -0.3 true cm}
\end{tabular}
\label{tab:ratio-RS-RY}
\end{center}
\end{table}

Thus, we understand the results by Angulo+ and Tumino+ on $R_S$ 
and $R_Y$ are significantly different from the others.
We expect more precise future observation of the $p$-wave $S(E)$ 
factors for $E \lesssim 100$ keV.

Other interesting  $dd\mu$ fusion experiments concerning the charge symmetry were given by Balin {\it et al.}~\cite{Balin1990}, 
Petitjean {\it et al.}~\cite{Petitjean1999}, 
and Balin {\it et al.}~\cite{Balin2011}. They found a temperature
dependence of $R_Y$, which gradually decreases from $R_Y \simeq 1.4$ at room temperature to \mbox{$R_Y \simeq 1.0$} at $T \lesssim 70$ K (cf. Fig.~10~\cite{Balin1990}, Fig.~3~\cite{Petitjean1999}, and Fig.~17~\cite{Balin2011}).
They explained it as, at room temperature, the $(dd\mu)_{J=v=1}$ state is formed resonantly by the Vesman's mechanism~\cite{Vesman1967}, and fusion takes place from the  $p$ wave of the $d$-$d$ system, whereas the non-resonant mechanism should dominate in generating the $(dd\mu)_{J=0}$ state at $T \lesssim 70$ K and fusion occurs in the \mbox{$s$ wave} of the $d$-$d$ system.
Note that, for the $s$-wave $d$-$d$ fusion reaction, $R_S \simeq 1.0$ was given in Refs.~\cite{Adya1981,Hale1990,Angulo1998,Nebia2002,
Tumino2014,Solovyev2024}.

\section{Summary}

The muon-catalyzed fusion ($\mu$CF) in the $dd\mu$ molecule, via reactions 
(1.3) and (1.4), was studied using the optical-potential model, and the tractable 
\mbox{$T$-matrix} model~\cite{Wu2024} that was proposed for studying the $dt\mu$ 
fusion and well approximates the elaborate coupled-channel framework by one of 
the authors (M.K.) and his collaborators~\cite{Kamimura2023}.
Our study is based on the use of the nuclear interactions that reproduce five 
cases of the $p$-wave astrophysical $S(E)$ factors of the reaction 
\mbox{$d + d \to\!^3{\rm He} + n$ or $t + p$}, in a broad energy region 
\mbox{$E \simeq$ 1 keV} to 
1 MeV~\cite{Angulo1998,Nebia2002,Arai2011,Solovyev2024,Tumino2014}
 (Fig.~\ref{fig:8line-sfactor}). None of these $S(E)$ factors has ever 
been used for studying the $d$-$d$ $\mu$CF.  

Since the nuclear interactions are phenomenological, we employed many sets of 
their parameters (Tables I and IV) to reproduce the $S(E)$ factors, and 
we demonstrated that the calculated results for the $dd\mu$ fusion were 
consistent among the parameter sets. 
Unfortunately, however, the five cases of $S(E)$ factors  themselves are 
significantly different from each other (Fig.~\ref{fig:8line-sfactor}), 
and the calculated results for some quantities show inconsistency.

Major conclusions are summarized as follows:

1) 
We calculated the fusion rate of $dd\mu$ molecule via
three methods: 
i) optical-potential model (Sec.~II), 
ii) \mbox{$T$-matrix} model calculation performed on channels 
5 and 8 in Fig.~\ref{fig:3body-jacobi} (Sec.~IV),
and iii) that on channels 4 and 7 (Sec.~VI). 
The calculated fusion rates of the $(dd\mu)_{J=v=1}$ state, $\lambda_{11}$,
are consistent with each other among these three methods 
and are summarized in Table~\ref{tab:lambda-muon-spec}.
However, depending on the five cases of the $p$-wave $S(E)$ factors, 
the fusion rates spread in a range 
$(1.8 - 5.1) \times 10^8 {\rm s}^{-1}$ 
which corresponds to {\rm effective} fusion rates
$(2.0 - 5.3) \times 10^8 {\rm s}^{-1}$, though including the observed 
values (cf: Table~\ref{tab:lambda-ref}).
Our fusion rate $\lambda_{11}$ supports the calculated 
literature values (cf.~Table~\ref{tab:lambda-ref})
which were derived using the $S(E \to 0)$ factor observed 
by Ref.~\cite{Adya1981}.

2)
Furthermore, we computed the branching ratio $R_Y$, Eq.~(8.2), 
of the $dd\mu$ fusion (1.3) and (1.4), 
employing the five cases of the $p$-wave $S(E)$ factors;
note that Bogdanova {\it et al.}~\cite{Bogdanova1982} 
pointed out $R_Y=1.46$ using the observed $S(E \to 0)$ factor 
of Ref.~\cite{Adya1981}.
Our ratio $R_Y$ ranges from $1.3$ to $1.5$  
when using the three $S(E)$ factors 
from Refs.~\cite{Nebia2002,Arai2011,Solovyev2024}(Table~\ref{tab:ratio-RS-RY}),  
which is consistent with 
the latest observed value $R_Y=1.455 (11)$~\cite{Balin2011}.
This indicates significant charge symmetry violation in the above 
reactions at low energies.  
Quite differently, $R_Y \!\approx \!1.0$ was obtained when using 
the two $S(E)$ factors from Refs.~\cite{Angulo1998,Tumino2014}. 
Check of these results require more precise observation (analysis) of 
the \mbox{$p$-wave} $S(E)$ factors of the reactions (1.1) and (1.2);

3)
The initial muon sticking probability $\omega_{\rm d}^{11}$ of the 
$(dd\mu)_{J=v=1}$ state was calculated with the definition
of Eq.~(5.1), using the absolute values of the transition rates to 
the $^3{\rm He}$-$\mu$ continuum and bound states ({Fig.~\ref{fig:rL-J1V1}).
We obtained $\omega_{\rm d}^{11}=0.133 \pm 0.001$ (Table~\ref{tab:vdtanVI}),
which agrees with the literature 
values $(0.131 - 0.134)$~\cite{Bogdanova1985,Hu1987,Haywood1991} 
based on the sudden approximation.
This is reasonable because the nuclear interaction in the $p$-wave $dd\mu$ system
is much smaller than that in the $s$-wave $dt\mu$ system.  
The present $\omega_{\rm d}^{11}$, after transformed to the effective
sticking probability $\omega_d^{\rm eff}({\rm th})$, agrees with the observed
sticking probability $\omega_d^{\rm eff}({\rm exp})=0.1224(6)$ 
within the error bars (Sec.~V);

4)
The momentum and energy spectra of the muon emitted by the $d$-$d$ $\mu$CF
(Figs.~\ref{fig:rK} and \ref{fig:rE}) were calculated for the first time.
The peak energies are located at 1.0 keV, much lower than the average energy 
of 8.2 keV, which is independent of the nuclear interactions and $S(E)$ factors.
This result will be helpful for future experiments that generate an ultra-slow 
negative muon beam using $d$-$d$ $\mu$CF for various 
\mbox{applications~\cite{Strasser1993,Strasser2000,Yamashita2021,Okutsu2021}.}  


\section*{Acknowledgements}

The authors would like to thank Prof.~Y.~Kino and Dr.~T.~Yamashita for 
valuable discussions. 
Thanks are also to Prof. A.~Tumino for providing the numerical data for 
the observed $S(E)$ factors of the $d$-$d$ reactions.
We are grateful to Prof. K.~Arai for providing the numerical results of the
four-nucleon calculation of the $d$-$d$ reaction 
and useful discussions on the results.
We would like to thank Prof. P. Descouvemont for helpful discussions on the 
experiments and analysis of the $d$-$d$ reactions. 
Thanks are also to Prof. A. Solovyev for providing the numerical results
of his microscopic cluster-model calculation. 
We are grateful to Prof. M.~Sato
for valuable discussions on the application of the neutrons generated by 
the $dd\mu$ fusion~\cite{Iiyoshi2023} and to Prof. \mbox{Y.~Nagatani}
for helpful discussions on the application of the muons emitted from the 
$dd\mu$ fusion.

This work is supported by
the Grant-in-Aid for Scientific Research on Innovative Areas,
``Toward new frontiers: Encounter and synergy of state-of-the-art
astronomical detectors and exotic quantum beam'', 
JSPS KAKENHI Grant Number JP18H05461.
This work is also supported by
Natural Science Foundation of Jiangsu Province (Grant No. BK20220122); 
National Natural Science Foundation of China (Grant No. 12233002); 
China Postdoctoral Science Foundation (Grant No. 2024M751369); 
and Jiangsu Funding Program for Excellent Postdoctoral Talent.



\end{document}